\documentclass[sigconf]{acmart}
\usepackage{natbib}
\usepackage{amsmath}
\usepackage{stfloats}

\title{\textit{OceanChat}: The Effect of Virtual Conversational AI Agents on Sustainable Attitude and Behavior Change \\ \vspace{0.2cm} \large{\textbf{Preprint}}}

\author{Pat Pataranutaporn}
\authornote{Both authors contributed equally to this research.}
\email{patppat@mit.edu}
\affiliation{
  \institution{MIT Media Lab}
  \institution{Massachusetts Institute of Technology (MIT)}
  \city{Cambridge}
  \state{MA}
  \country{USA}
}

\author{Alexander Doudkin}
\authornotemark[1]
\email{doudkin@mit.edu}
\affiliation{
  \institution{MIT Media Lab}
  \institution{Massachusetts Institute of Technology (MIT)}
  \city{Cambridge}
  \state{MA}
  \country{USA}
}

\author{Pattie Maes}
\email{pattie@media.mit.edu}
\affiliation{
  \institution{MIT Media Lab}
  \institution{Massachusetts Institute of Technology (MIT)}
  \city{Cambridge}
  \state{MA}
  \country{USA}
}

\begin{teaserfigure}
  \centering
  \includegraphics[width=\textwidth]{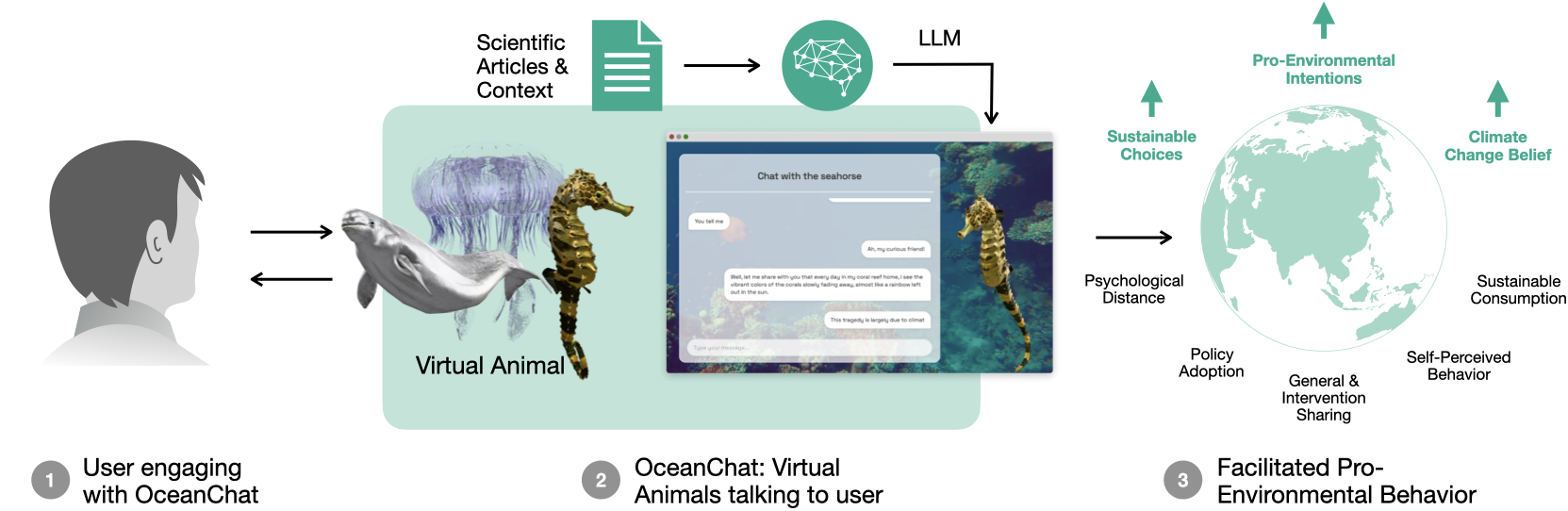}
  \caption{Explanatory Overview of the OceanChat Intervention}
  \label{fig:overview}
\end{teaserfigure}

\begin{abstract}
Marine ecosystems face unprecedented threats from climate change and plastic pollution, yet traditional environmental education often struggles to translate awareness into sustained behavioral change. This paper presents \textit{OceanChat}, an interactive system leveraging large language models to create conversational AI agents  represented as animated marine creatures—specifically a beluga whale, a jellyfish, and a seahorse—designed to promote environmental behavior (PEB) and foster awareness through personalized dialogue. Through a between-subjects experiment (N=900), we compared three conditions: (1) Static Scientific Information, providing conventional environmental education through text and images; (2) Static Character Narrative, featuring first-person storytelling from 3D-rendered marine creatures; and (3) Conversational Character Narrative, enabling real-time dialogue with AI-powered marine characters. Our analysis revealed that the Conversational Character Narrative condition significantly increased behavioral intentions and sustainable choice preferences compared to static approaches. The beluga whale character demonstrated consistently stronger emotional engagement across multiple measures, including perceived anthropomorphism and empathy. However, impacts on deeper measures like climate policy support and psychological distance were limited, highlighting the complexity of shifting entrenched beliefs. Our work extends research on sustainability interfaces facilitating PEB and offers design principles for creating emotionally resonant, context-aware AI characters. By balancing anthropomorphism with species authenticity, \textit{OceanChat} demonstrates how interactive narratives can bridge the gap between environmental knowledge and real-world behavior change.
\end{abstract}


\setcopyright{none}
\renewcommand\footnotetextcopyrightpermission[1]{}
\renewcommand\footnotetextcopyrightpermission[1]{} 
\begin{document}
\maketitle

\section{Safe and Responsible Innovation Statement}
The OceanChat project is committed to ethical and responsible innovation, aligning with ICMI 2025’s theme of Safe and Responsible Multimodal Interaction. We address societal impacts by fostering pro-environmental behavior through AI-driven marine characters, promoting inclusivity via accessible web platforms. Ethical considerations include transparent AI use, informed consent, and anonymized data to ensure privacy. We mitigate bias by randomizing species and narratives, balancing anthropomorphism with ecological authenticity to avoid misrepresentation. Potential risks, such as over-reliance on AI for environmental education, are addressed by emphasizing human-nature connections.

\section{Introduction}
Climate change and its consequences represent the defining challenge of our time, threatening ecosystems, communities, and the very stability of Earth's life-supporting systems. Despite mounting scientific evidence and growing public awareness, the gap between environmental knowledge and action remains a persistent challenge in addressing climate change \cite{Knutti2019}.

Dava Newman, former Deputy Administrator of NASA, once remarked that Earth needs ambassadors—voices that can speak for our planet and its diverse inhabitants in the face of unprecedented environmental challenges. This provocative idea takes on new relevance in our age of AI. What if we could create Earth ambassadors that give voice to non-human perspectives?

The world's oceans face an unprecedented crisis as plastic pollution continues to surge, threatening marine ecosystems on a global scale  \cite{Geyer2017}. Recent quantitative assessments reveal that by 2022, approximately 139 million tonnes of plastic have accumulated in aquatic ecosystems worldwide, with documented adverse effects on marine biodiversity, including reproductive disruption in aquatic species and agricultural impacts \cite{Cole2011, Meszaros2023, OECD2022a, OECD2022b}. Furthermore, emerging research has identified microplastic particles in human circulatory systems, correlating with elevated risks of cardiovascular events and other physiological complications \cite{Marfella2024}. 

Despite extensive literature on sustainability messaging and pro-environmental behavior (PEB) interventions, some conventional methodologies demonstrate first promising results, while others are still limited in their effectiveness - especially in the long-term \cite{Hadler2022, Nguyen2019, Cavaliere2020}. Traditional environmental education methods often struggle to evoke the emotional resonance required to catalyze meaningful, long-term behavioral change \cite{Markowitz2018}.

Recent advances in artificial intelligence, particularly in large language models (LLMs) and conversational AI, present new opportunities for addressing these challenges. These technologies offer the potential to create dynamic, personalized environmental narratives that might help bridge the gap between human experience and marine conservation. However, they also raise important questions about representation, authenticity, and the ethical implications of simulating non-human perspectives.

Drawing on Haraway's seminal work on multispecies relationships and her concept of "making kin" with non-human species, we explore how technological interventions can reshape human-ocean connections  \cite{Haraway2016}. Haraway's framework of companion species offers valuable insights into how humans might develop more meaningful relationships with marine ecosystems through technological mediation. Building on these theoretical foundations, we present \textit{OceanChat}, an interactive system featuring AI-generated marine characters designed to foster environmental awareness and sustainable behaviors. By providing engaging, character-driven conversations, \textit{OceanChat} seeks to bridge the gap between environmental awareness and pro-environmental behavior. While acknowledging that AI cannot truly capture or represent marine life perspectives, we investigate how these characters might serve as Newman's envisioned Earth ambassadors, giving voice to marine species in the crucial work of environmental education and advocacy.

This approach builds on recent HCI research examining how conversational agents can support environmental engagement \cite{Giudici2024, Hillebrand2021, Breiter2024} while extending it through the integration of advanced conversational AI and species-specific character design. Prior work has explored various approaches to environmental communication through technology, from eco-feedback systems providing real-time energy consumption data to gamified sustainability interfaces encouraging pro-environmental behaviors \cite{Strengers2011, Hillebrand2021}. Research has also examined the role of emotional connection in environmental advocacy and the use of anthropomorphism in conservation messaging \cite{Nabi2018, RootBernstein2013}. Interactive technologies have shown promise in creating more engaging environmental experiences, as demonstrated by virtual reality applications for ocean acidification awareness and mobile apps for citizen science initiatives \cite{Fauville2021, Johnson2020}.

However, while these studies have advanced our understanding of environmental communication through technology, there remains a critical gap in examining how scalable solutions incorporating advanced conversational AI and species-specific character design can create more personalized and emotionally resonant environmental engagement experiences. Most existing approaches rely on static information delivery, specialized hardware, or limited interactivity, potentially missing opportunities for widespread, accessible engagement through mobile and web platforms. Building on these theoretical foundations and technological capabilities, we conducted a mixed-methods study examining how interactions with three different marine characters—a beluga whale, jellyfish, and seahorse—influenced participants' environmental attitudes and actions. These species were selected to represent diverse forms of marine life with varying degrees of charisma and anthropomorphic potential. Our experimental design compared three conditions: (1) traditional scientific narratives with static marine life images, representing conventional environmental education approaches, (2) first-person narratives from marine creatures with virtual characters but no interaction, exploring the impact of character-driven storytelling, and (3) \textit{OceanChat}'s interactive AI-generated marine characters with conversational capabilities, investigating the potential of dynamic, engaging communication efforts.

In alignment with our sustainability ethos, we critically assessed the environmental impact of our intervention, ensuring that its implementation adheres to responsible AI practices. This research makes three primary contributions to the fields of human-computer interaction, behavioral psychology, and environmental communication:
\begin{enumerate}
    \item \textbf{Empirical Insights into Character Design and Pro-Environmental Behavior} \\
    Our study provides robust quantitative and qualitative evidence (N=900) on how marine species and interaction modalities influence environmental attitudes, intentions, and behaviors. Key findings reveal that character-driven narratives—particularly those utilizing interactive conversational agents—are effective in fostering pro-environmental engagement, though their impact varies across species and perceptual dimensions. The study highlights the role of factors such as perceived anthropomorphism, animacy, and empathy in mediating these effects, offering a nuanced understanding of the interplay between design elements and aspects of pro-environmental behavior.
    \item \textbf{Design Principles for AI-Generated Environmental Characters} \\
    We propose actionable design principles for creating effective AI-generated characters for environmental education. These principles emphasize balancing anthropomorphic features with authentic species representation to maximize emotional connection while maintaining ecological credibility. The concept of “graduated anthropomorphism” is introduced, recommending context-specific adjustments in character design to align with interaction goals. Additionally, we outline strategies for leveraging emotional scaffolding and contextual resonance to create immersive, persuasive interactions.
    \item \textbf{Synhetic pre-evaluation of Study} \\
    \textit{OceanChat} underwent a synthetic pre-evaluation—a novel approach to leverage large language models (LLMs) to simulate diverse participant interactions and provide detailed feedback before deploying the system to human users. By integrating synthetic reviewers, the study introduces a rigorous pre-testing layer that enhances the reliability and validity of the final intervention, setting a precedent for incorporating LLM-driven pre-evaluation in behavioral and sustainability research.
\end{enumerate}

Our research extends HCI work on persuasive interfaces for pro-environmental behavior while providing practical guidelines for incorporating AI-generated characters in environmental education. More broadly, it contributes to ongoing discussions about the role of AI in mediating human relationships with the natural world, offering insights into how technological interventions might support rather than supplant direct environmental connection and responsibility.

The following sections detail related work, our methodology, findings, and implications for design. We begin with a more detailed review of related work, followed by a description of the \textit{OceanChat} system and our experimental design. We then present our results, focusing on the differential effects of species selection and interaction modality on environmental engagement. Finally, we discuss the implications of our findings for the design of AI-based environmental communication systems and suggest directions for future research in this emerging field.

\section{Related Work}
As sustainability challenges become increasingly urgent, researchers have turned to interactive AI systems—especially those powered by large language models (LLMs)—as potentially powerful drivers of pro-environmental behavior. Below, we situate our work within two key areas of prior research: the emerging role of LLMs in persuasive interactions for (pro-environmental) behavior change and methodological considerations that incorporate both real and synthetic participants.

\subsection{LLMs and Persuasive Interactions for Sustainability}
The rapid growth of large language models (LLMs) has not only transformed how individuals interact with AI—exemplified by platforms like OpenAI’s ChatGPT—but has also opened new avenues for influencing environmental knowledge, attitudes, and behaviors. By leveraging principles from behavioral science and persuasive design, these conversational systems can adapt messages in real-time to address individual misconceptions and motivations through personalized interactions \cite{Fogg2003, Giudici2024, Giudici2024b}. In parallel, advancements in behavioral psychology have revealed promising strategies for fostering sustainable behavior. Psychological methods such as leveraging dynamic social norms, goal setting, and emotionally resonant storytelling have demonstrated the potential to motivate individuals towards adopting pro-environmental habits \cite{Breiter2024, Vlasceanu2024, Gifford2011a, Aavik2022UsingChoices}. When integrated into AI-driven interfaces, these approaches can be deployed at scale to offer tailored suggestions that strengthen users’ sense of self-efficacy and commitment to environmental goals, thereby contributing to broader sustainability efforts.

Empirical research increasingly demonstrates these systems’ ability to shift beliefs across diverse contexts. For instance, LLM-powered agents have been shown to counteract misinformation and conspiracy theories, guide users toward healthier behaviors, and enhance education effectiveness \cite{Wang2023, Aggarwal2023, Costello2024}. Through iterative dialogue, chatbots can tailor feedback to users’ specific barriers—such as time constraints or skepticism—thereby enhancing both the persuasiveness and relevance of pro-environmental messages \cite{Aggarwal2023, Giudici2024b}. 

Building on these insights, \textit{OceanChat} explores how LLM-driven conversations might create richer, more empathic connections with marine ecosystems than static or one-way interventions. While prior studies confirm the efficacy of conversational agents for changing attitudes, less is known about how fully \textit{embodied}, character-based narratives might deepen engagement with—and commitment to—sustainability goals.

Recent experiments underscore the importance of multi-strategy approaches. For instance, Costello et al. found that short, tailored interventions generated through GPT-4 Turbo significantly diminished conspiracy-theory beliefs over time, illustrating how fact-based appeals can complement empathy-building dialogues \cite{Costello2024}. Meanwhile, Giudici’s chatbot delivering energy-saving tips combined multiple persuasive cues (e.g., social comparisons, immediate feedback) to improve user engagement, though the translation to actual behavior varied \cite{Giudici2024}.

Taken together, these findings suggest that effective sustainability interventions often blend factual content with emotional resonance and interactive affordances. In the marine conservation domain, such hybrid approaches could be particularly beneficial, as users’ emotional connection to non-human life forms is a key driver of behavior change. With \textit{OceanChat}, we extend this line of inquiry by using first-person, AI-generated marine characters to evoke empathy, frame scientific consensus on climate impacts, and deliver context-aware guidance on sustainability practices.

\subsection{Methodological Considerations: Real and Synthetic Participants}
Designing and testing novel AI interventions at scale raises unique methodological questions. While human-subject experiments remain the gold standard for measuring authentic behavioral and attitudinal change, LLM-based synthetic pilot studies have emerged as a complementary technique \cite{SyntheticStudy2023, Long2024}. By simulating participants from different demographic backgrounds, researchers can pre-test interventions for potential flaws, refine prompts for clarity, and evaluate initial user flows before deploying a system to a larger pool of human participants.

We adopt this dual approach, integrating a synthetic pilot study with our eventual human-subject trial. In doing so, we first tested \textit{OceanChat}’s conversational prompts, ensuring the characters’ personalities, narrative tone, and factual accuracy met our design goals. Synthetic participants flagged potential pitfalls—such as confusing transitions or underexplained technical terms—that might reduce the intervention’s effectiveness. After refining these elements, we proceeded to test the system with real participants, allowing us to focus on genuine emotional engagement and reported behavior change rather than debugging user-flow issues.

By combining LLM-based research techniques, best practices in persuasive design, and real-user testing, \textit{OceanChat} investigates how conversational AI agents can drive deeper, more lasting pro-environmental attitudes and behaviors. This dual-method strategy ensures robust study design while capitalizing on the rapid iteration possible through LLM-driven synthetic pilots.

\section{Methodology}
\begin{figure*}[!b] % '!b' tries to force bottom
    \centering
    \includegraphics[width=1\linewidth]{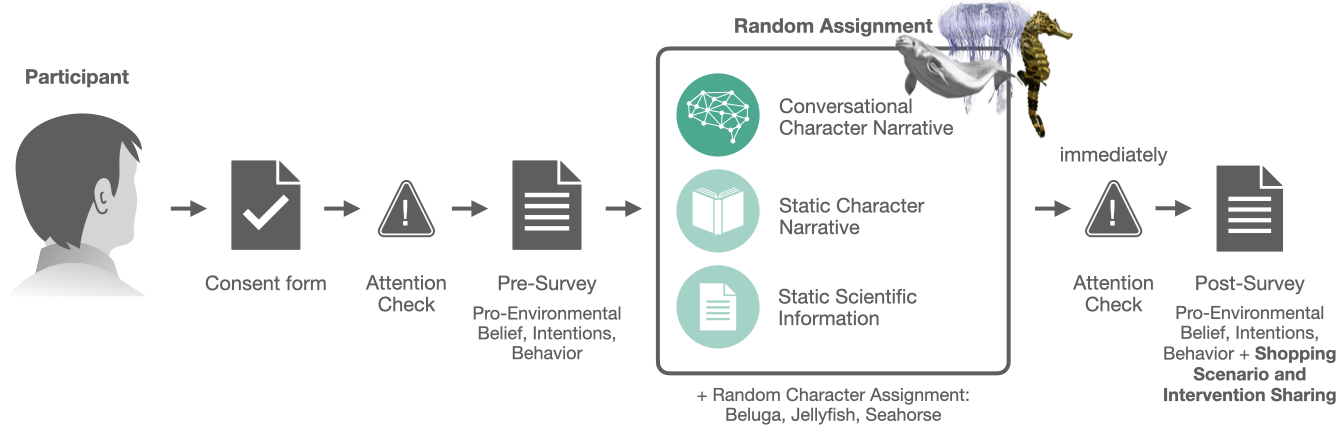}
    \caption{Study Design Overview}
    \label{fig:ocean-chat-study-design}
\end{figure*}

The methodology of this study was designed to explore the potential of AI-driven conversational agents in fostering pro-environmental attitudes and behaviors. By combining principles from behavioral psychology, environmental science, and human-computer interaction, the study operationalizes a robust framework for evaluating the effectiveness of \textit{OceanChat}. This section outlines the experimental conditions, data collection methods, and analytical strategies employed to ensure rigorous and replicable findings. Our research question specifically focuses on comparing the effectiveness of interactive AI agents versus static methods in communicating about climate change. The methodology addresses this question through randomized controlled experiments that isolate the effects of interactivity while maintaining consistent environmental messaging across all participant groups.
\subsection{Study Design} 
This paper is using a between-subjects design with three experimental conditions:
\begin{itemize}
    \item \textbf{"Static Scientific Information"}: A text-based intervention presenting a succinct scientific excerpt and a supporting image of a marine animal.
    \item \textbf{"Static Character Narrative"}: A text-based intervention featuring a personal story from the animals perspective, accompanied by an animated 3D model.
    \item \textbf{"Conversational Character Narrative"}: An interactive, chat-based intervention that provides a personal animal narrative, coupled with an animated 3D model offering real-time dialogue.
\end{itemize}

Within each of these three conditions, the study incorporates an additional layer of even randomization by featuring one of three maritime characters: Three-Spot seahorses, Moon Jellyfish, or Beluga Whales. These species were chosen because they are either endangered or significantly affected by climate change and plastic pollution \cite{Rapp2021, Williams2021, Dong2018}. This nested randomization allows researchers to examine whether the effectiveness of each presentation format varies depending on the featured species.

This comprehensive design enables us to investigate multiple factors: how different presentation formats affect engagement and learning, whether certain marine species resonate more strongly with participants, and if there are any interesting interactions between the presentation format and the featured species. The between-subjects approach ensures clean data collection without the confounding effects that might arise from participants experiencing multiple conditions.

\subsection{Chatbot Design}
The \textit{OceanChat} chatbot design combines behavioral psychology principles with cutting-edge technical implementation to create engaging, character-driven interactions that promote pro-environmental behavior.
\subsubsection{Behavioral Design}
In \textit{OceanChat}, each interaction with a marine character begins with the character narrating a short, vivid story about their life, challenges, and the impacts of plastic pollution or climate change. This story is generated by a LLM and tailored to the specific character, such as “Luno,” a male moon jellyfish.

Following the story, the user engages in a three-round conversation inspired by the methodology from \citeauthor{Costello2024} persuasive dialogue \cite{Costello2024}. Each response by the character integrates the user’s input, builds on their message with oceanic anecdotes, and uses behavioral psychology principles to nudge pro-environmental behavior. Responses are concise (2–3 sentences) and end with a reflective question to sustain engagement. After the third user input, the conversation concludes with a brief, urgent, and hopeful call to action, emphasizing marine protection and sustainable practices. This structured approach ensures meaningful yet time-efficient interactions.

\begin{figure*}[htbp]
  \centering
  \includegraphics[width=\linewidth]{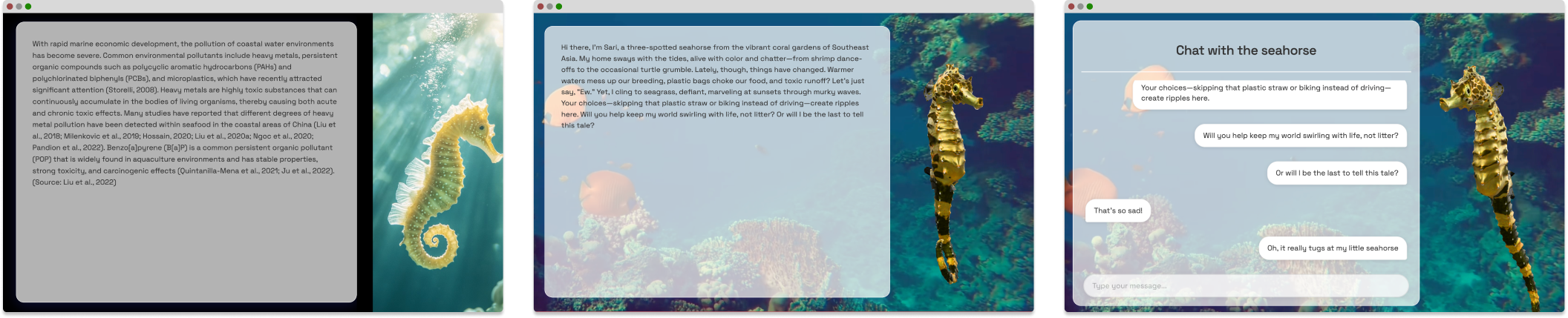}
  \par\vspace{0.2cm}
  \includegraphics[width=\linewidth]{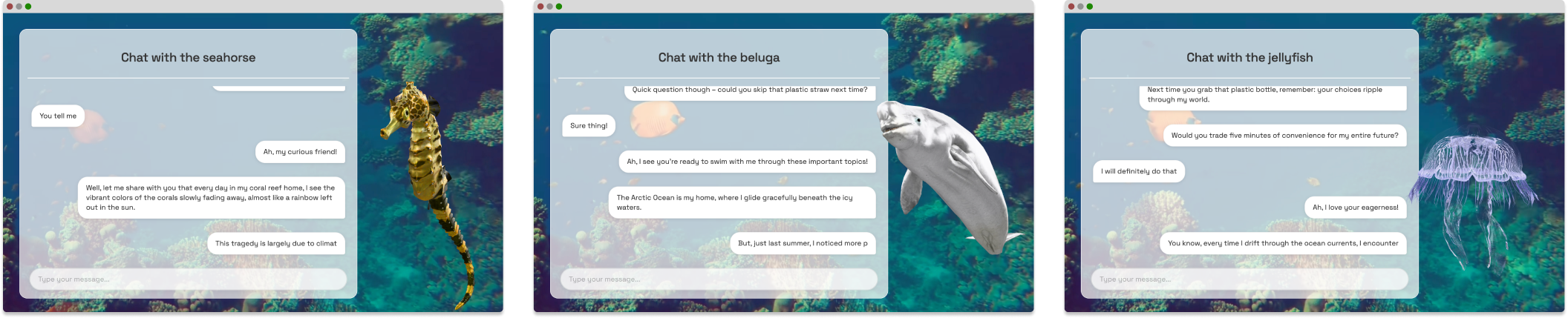}
  \caption{Overview of all three conditions for the seahorse character (top) and OceanChat interfaces of all three animals (bottom)}
  \label{fig:interfaces}
\end{figure*}

\subsubsection{Technical Implementation}

\textit{OceanChat} was developed to deliver contextually relevant, character-specific dialogue using OpenAI's \texttt{gpt-4o-mini} model \cite{openai2025}. This model was configured to produce responses aligning with each of the study conditions, with content tailored to the chosen marine character. To enhance user immersion, the system incorporated OpenAI's Text-to-Speech (TTS) model \texttt{tts-1} that converted AI-generated text into realistic audio output—with each character maintaining a different voice \cite{openai2025}. 

A \texttt{vue-js} frontend, hosted on Replit, presented the Chat Interface which managed user input, displayed real-time chat responses, and streamed corresponding audio \cite{replit, vuejs}. All experimental conditions—static information, static narrative, and AI-driven conversational narrative—were implemented within a single, unified interface, thus ensuring consistent visual design and interaction across conditions.

In the character conditions, the animation controller managed a comprehensive library of character-specific behaviors, including ambient movements, conversational gestures, and emotional expressions. Three-dimensional marine life models were rendered through Google's \texttt{model-viewer}, which provided hardware-accelerated WebGL rendering and sophisticated materials handling \cite{google2025}. The viewer was extended with custom shaders to achieve realistic underwater effects. The interface architecture maintained strict separation between the presentation layer and the underlying conversation logic. Real-time chat responses were processed through an asynchronous pipeline that coordinated text generation, speech synthesis, and animation triggering. 

\subsection{Ethical Considerations}
This study adhered to ethical guidelines, ensuring participant welfare, data security, and transparency in AI use. Informed consent was obtained, with participants fully briefed on the study’s objectives, procedures, and their right to withdraw at any time. The result data was anonymized and securely stored, ensuring participant confidentiality.

Participants were explicitly informed of the AI-based nature of the conversational agents, and emotional risks were minimized by avoiding sensitive content. In addition, the study evaluated the environmental impact of the AI system, optimizing computational resources to minimize its carbon footprint.

\subsection{Sample Size}
To ensure sufficient statistical power, we conducted an \textit{a priori} power analysis using G*Power 3.1.9.7 \cite{Erdfelder2009}. Our study employed a 3 (\textit{Intervention Type}) $\times$ 3 (\textit{Animal Condition}) between-subjects design (nine groups total). We aimed to detect a medium effect size ($f = 0.25$) at an $\alpha$ level of 0.05 with $1-\beta=0.80$ power for the main effects and interactions.

Using the ``fixed effects, special, main effects and interactions'' module in G*Power, the analysis indicated that a minimum of 270 participants (i.e., 30 per group) would be required to detect an overall medium effect with adequate power. However, we increased our target sample size to 900 participants for three reasons:
\begin{enumerate}
    \item We anticipated potential attrition due to participant non-compliance, attention-check failures, or technical issues.
    \item We expected to run multiple comparisons across a range of outcome variables and sought to maintain acceptable statistical power after any adjustments or corrections.
    \item We planned for potential subgroup and exploratory analyses (e.g., examining interactions between \textit{Intervention Type} and \textit{Animal Condition}).
\end{enumerate}

\subsection{Participant Recruitment}
This study was preregistered on AsPredicted (Registration [retracted for anonymity]). Data collection was conducted through Cloud Research with automated balancing features enabled. In total, we recruited all 900 participants through Cloud Research. After excluding participants who failed attention checks, completed the survey too quickly (under 3 minutes), or provided incomplete data, the final analytic sample was randomly evened out and ranged from $n=683$ to $n=782$ across the various outcome measures.

\subsection{Statistical Analysis and Model Specification}
Our primary analysis employed a comprehensive statistical approach to examine the effects of our intervention on multiple environmental and perceptual outcomes as we conducted a series of linear regression analyses.

\subsubsection{Dependent Variables}
To measure broadly and understand in which dimension of pro-environmental behavior and perception \textit{OceanChat} can have an impact, the following dimensions were measured. 

\begin{itemize}
    \item \textbf{Environmental Beliefs}
    \begin{itemize}
        \item \textbf{Climate Change Belief (Discrete 0--100):} Based on validated survey tournament scale, measured both pre- and post-intervention \cite{Vlasceanu2024}.
        \item \textbf{Psychological Distance (Discrete 1--7):} Validated distance scale, encompassing temporal and spatial subscales \cite{vanV2021}.
        \item \textbf{Climate Policy Adoption (Discrete 0--100):} Based on validated survey tournament scale, measured both pre- and post-intervention \cite{Vlasceanu2024}.
    \end{itemize}

    \item \textbf{Behavioral Measures}
    \begin{itemize}
        \item \textbf{Pro-environmental Behavior (Discrete 1-5):} A custom experimental metric quantifying self-reported single-use plastic consumption.
        \item \textbf{Pro-environmental Intentions (Discrete 1-5):} Two custom experimental metrics assessing likelihood of choosing refillable products and intended reduction of single-use plastic.
        \item \textbf{Sustainable Consumption (Discrete 1-7):} GREEN scale administered pre- and post-intervention \cite{Haws2014}.
        \item \textbf{Sustainable Choice Preferences (post-intervention only):} An experimental product selection task (6 sustainable vs. 6 non-sustainable) for calculating a sustainable product ratio.
    \end{itemize}

    \item \textbf{Agent Perception Measures (Discrete 1-5, all post-in\-ter\-vention only):} Measurement instrument for perceived robot traits ("Godspeed" scale) \cite{Bartneck2009}.
    \begin{itemize}
    \item \textbf{Perceived Anthropomorphism:} The degree to which participants attribute human-like characteristics, behaviors, or mental states to the virtual character's personality and actions.
    
    \item \textbf{Perceived Animacy:} The extent to which participants view the virtual character as possessing life-like qualities through its movements, behaviors, and responses within the virtual environment.
    
    \item \textbf{Perceived Likeability:} The overall positive emotional response and appeal that participants experience when interacting with the virtual character.
    
    \item \textbf{Perceived Intelligence:} The degree to which participants attribute cognitive capabilities and reasoning abilities to the virtual character based on its interactions and responses.
    
    \item \textbf{Perceived Safety:} The extent to which participants feel psychologically secure and comfortable during their interactions with the virtual character.
    
    \item \textbf{Empathy with the Animal (experimental addition):} The degree to which participants form emotional connections and understanding with the virtual animal character's portrayed experiences and states.
    
    \item \textbf{Climate Change Impact on Animal (experimental addition):} The level of participant awareness and concern regarding how climate change affects the real animal species that the virtual character represents.
    \end{itemize}

    \item \textbf{Information Sharing Measures}
    \begin{itemize}
        \item \textbf{Willingness to Share Information (Binary):} Adapted from validated survey tournament scale, measured both pre- and post-intervention \cite{Vlasceanu2024}.
        \item \textbf{Intervention Sharing (Binary, post-intervention only):} An experimental measure indicating whether participants shared the intervention content.
    \end{itemize}
\end{itemize}

\begin{figure}[h]
    \centering
    \includegraphics[width=1\linewidth]{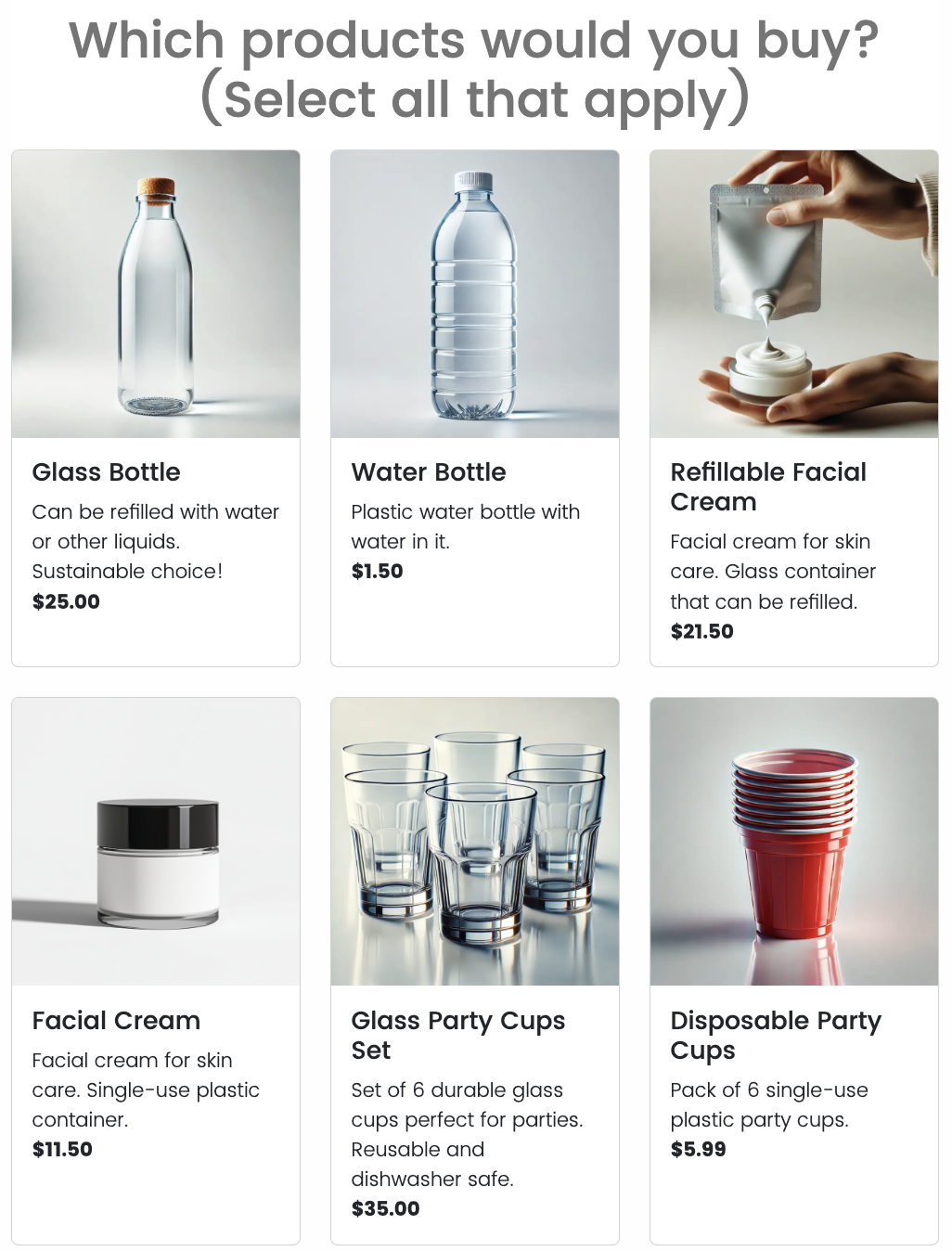}
    \caption{Excerpt of Post-Intervention Product Shopping Scenario (Sustainable Choice Preferences)}
    \label{fig:shopping_scenario}
\end{figure}

\subsubsection{Regression Model}

The general form of our regression equation is:
\begin{align*}
Y_{\text{post}} &= \beta_0 + \beta_1(\text{Experimental Condition}) + \beta_2(\text{Animal Condition}) \\
                &\quad + \beta_3(Y_{\text{pre}}) + \beta_4(\text{Demographics}) \\
                &\quad + \beta_5(\text{Perceived Variables}) + \epsilon
\end{align*}

Where $Y_{\text{post}}$ represents the post-intervention outcome measure, with Experimental Condition and Animal Condition serving as categorical variables representing the intervention conditions and marine species respectively. The models incorporated pre-intervention measures ($Y_{\text{pre}}$), demographic controls (age, gender, education, location, participant ID), and perceived variables (anthropomorphism, animacy, likeability, intelligence, safety, empathy with the animal, and climate change impact awareness).

For each outcome variable, we constructed regression models that controlled for pre-existing attitudes, demographic factors, and participant characteristics. All models included perceived variables as covariates, with bootstrap standard errors (1,000 replications) employed to ensure robust results. The analysis assessed three categories of outcome variables: environmental attitudes and beliefs, behavioral intentions and choices, and intervention effectiveness measures.

Model fit was evaluated using multiple indicators, including $R^2$ values, adjusted $R^2$ values, root mean square error (RMSE), and Wald chi-square statistics. To account for potential pre-existing differences, we included pre-intervention measures as control variables where available. The analysis used standardized coefficients to facilitate comparison across different outcome measures, with statistical significance assessed at conventional levels ($p < 0.05$) and marginal effects noted at $p < 0.10$.

To ensure robustness, we conducted sensitivity analyses with different model specifications and control variables. The results remained consistent across these alternative specifications, supporting the reliability of our findings. Participants who did not consent or failed attention checks were also excluded.

This comprehensive analytical approach allowed us to examine both direct and indirect effects of our intervention while controlling for relevant confounding variables and accounting for the nested structure of our experimental design.

\section{Results}
The results section presents a comprehensive analysis of how OceanChat's AI-generated marine characters influenced environmental attitudes and behaviors. We begin by examining the outcomes of our synthetic participant evaluation, which provided crucial insights for refining the system's conversational capabilities before deployment with human participants. The analysis then progresses through several interconnected dimensions of participant response.

Our primary analysis encompasses four key areas: First, we examine how different marine species representations and interaction modalities shaped participants' perceptions and emotional connections. Second, we analyze the intervention's impact on environmental beliefs and psychological distance toward climate change. Third, we investigate behavioral outcomes, including both self-reported actions and measurable choice preferences. Finally, we assess the system's effectiveness in promoting information sharing and sustained engagement with environmental issues.

Throughout these analyses, we employ robust statistical methods to untangle the complex relationships between character design, interaction modality, and environmental engagement. The findings reveal both promising avenues for AI-driven environmental communication and important limitations that warrant further consideration. 

\subsection{Synthetic Participants Evaluation}
Our evaluation of OceanChat began with a comprehensive synthetic testing phase aimed at validating and refining its conversational capabilities prior to deployment with human participants. This phase consisted of two stages: an interaction stage with simulated users and a subsequent review stage to assess conversation quality.

In the interaction stage, we deployed 100 large language model (LLM) agents, each prompted as a human with distinct conversational traits, to engage with OceanChat’s Conversational Narrative condition. These agents generated a substantial dataset of interactions, allowing us to evaluate the system’s ability to maintain consistent and meaningful dialogue across varied conversational paths. The dependent variable \textit{Sustainable Consumption} was simulated as the primary focus of this phase.

\begin{figure*}[ht]
    \centering
    \includegraphics[width=1\linewidth]{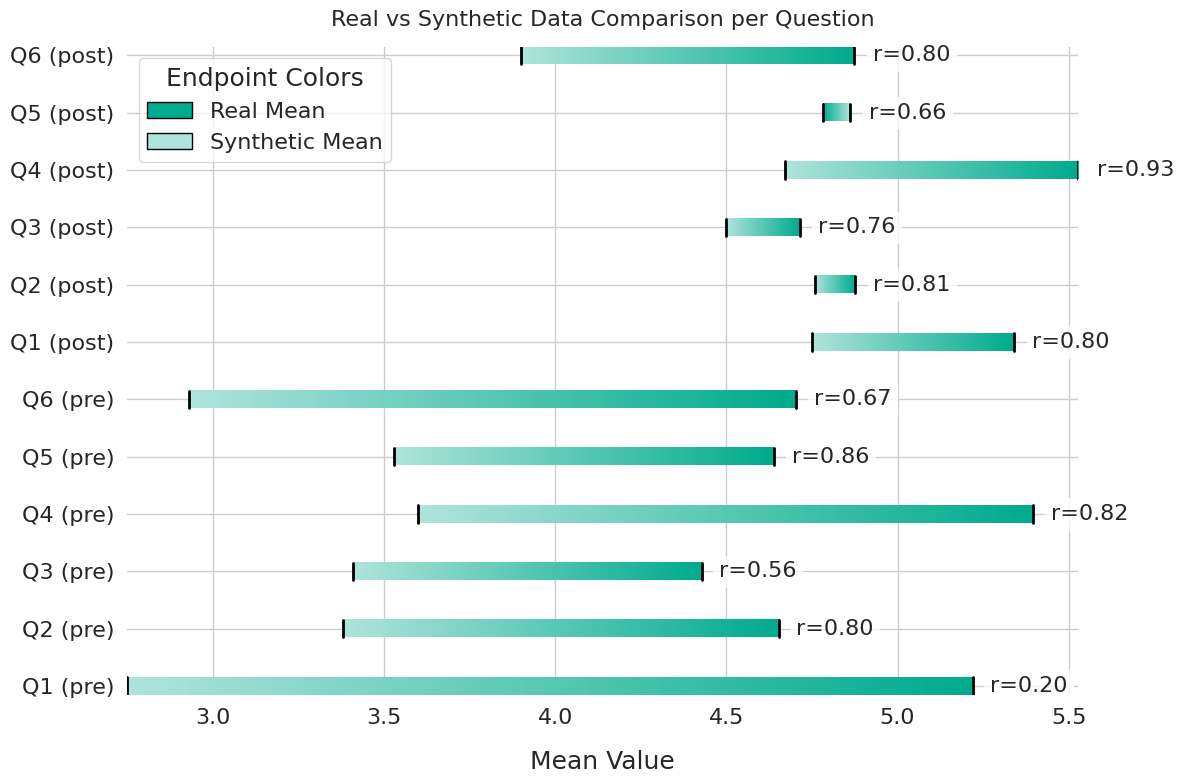}
    \caption{Comparison of synthetic and real participant results for the dependent variable "Sustainable Consumption".}
    \label{fig:synthetic_comparison}
\end{figure*}

The synthetic results, depicted in Figure~\ref{fig:synthetic_comparison}, reveal notable differences between the responses of synthetic agents and real participants. In the pre-intervention phase, the synthetic data produced a Kolmogorov–Smirnov (KS) statistic of 0.378133 ($p < 0.001$), indicating a statistically significant divergence from the real data. Post-intervention, the KS statistic decreased to 0.240819 ($p < 0.001$), suggesting a reduced, yet still meaningful, discrepancy. Moreover, the overall change from pre- to post-intervention yielded a KS statistic of 0.449195 ($p < 0.001$), highlighting the challenges inherent in fully replicating human responses with synthetic agents.

Complementing the KS analysis, we computed the Pearson correlation coefficients between the binned histogram densities of the real and synthetic responses for each question. These correlations ranged from approximately 0.20 to 0.93. Notably, for some questions the high correlations (up to 0.93) indicate that, despite significant KS differences, the synthetic responses capture the overall trend of the real data. Conversely, lower correlations (around 0.20) underscore substantial differences in distribution shape. Together, the KS statistics and correlation coefficients provide a nuanced view: while the KS test confirms statistically significant differences between the datasets, the degree of similarity—as measured by correlation—is highly question-dependent, informing both the limitations and the potential of synthetic data for refining prompting techniques.

Following the interaction stage, we conducted an extensive review process involving 300 additional LLM agents, tasked with evaluating various aspects of the conversations. These reviewers assessed emotional engagement, prompt adherence, motivational effectiveness, and behavioral impact. As shown in Table \ref{tab:means}, the reviewers reported stronger performance across all metrics when considering complete conversation threads (\textit{All Messages}) compared to isolated final responses (\textit{Last Message}). Notably, emotional engagement and motivational impact exhibited significant improvements in the \textit{All Messages} evaluation.

\begin{table}[ht]
    \centering
    \begin{tabular}{lcc}
        \toprule
        \textbf{Metric}            & \textbf{Last Message} & \textbf{All Messages} \\
        \midrule
        Emotion                    & 2.591667              & 4.141667              \\
        Prompt Adherence           & 2.983333              & 4.050000              \\
        Motivation                 & 3.362500              & 4.316667              \\
        Behavior                   & 2.566667              & 3.500000              \\
        \bottomrule
    \end{tabular}
    \caption{Mean evaluation scores for 100 conversations reviewed by 300 LLM agents.}
    \label{tab:means}
\end{table}

A detailed comparison of prompt variations, as summarized in Table \ref{tab:summaries}, provided further insights for system refinement. Feedback from the reviewers highlighted significant deficiencies in emotional depth, engagement, and storytelling quality in the \texttt{seahorse\_new} prompt, emphasizing the need for more vivid imagery, emotional clarity, and stronger narrative connections to human behavior. Similarly, the \texttt{seahorse\_og} prompt received criticism for inconsistent adherence to guidelines, a lack of urgency, and insufficient emotional resonance. These findings underscored the importance of enhancing narrative engagement and aligning prompt structure with the intended behavioral outcomes.

\begin{table}[ht]
    \centering
    \resizebox{0.5\textwidth}{!}{%
        \begin{tabular}{lp{10cm}}
            \toprule
            \textbf{Dataset}  & \textbf{Summary} \\
            \midrule
            \texttt{seahorse\_new} & The statements collectively highlight a significant deficiency in emotional depth, engagement, and storytelling quality. They emphasize the need for stronger narrative elements, clearer connections to human behavior, and better adherence to guidelines and prompts. Overall, the feedback suggests a lack of vivid imagery and emotional clarity, indicating that the content does not effectively resonate with or captivate its audience. \\[1ex]
            \texttt{seahorse\_og}  & The statements collectively highlight the need for improved emotional engagement and adherence to prompt guidelines in responses. Many responses were criticized for lacking urgency, emotional depth, and consistent compliance with the specified format. While some aspects of engagement were present, they often fell short, missing reflective questions and poetic elements. Overall, the agent needs to enhance emotional connection and follow prompt structure more closely. \\
            \bottomrule
        \end{tabular}
    }
    \caption{Summary feedback comparing two system prompts from 300 LLM review agents.}
    \label{tab:summaries}
\end{table}

This synthetic evaluation process proved instrumental in identifying and addressing potential interaction challenges prior to human deployment. It also provided quantitative benchmarks for assessing OceanChat’s conversational capabilities. The insights gained directly informed the refinement of OceanChat’s prompting strategies and interaction design, with a particular focus on improving storytelling quality and strengthening the connections between environmental messages and human behavior.

\subsection{Primary Analysis}
This section presents a comprehensive analysis of how different narrative approaches and marine species representations influence participants' PEB. The analysis is structured around our four key measurement areas: (1) Environmental Beliefs, encompassing climate change beliefs, psychological distance, and climate policy adoption; (2) Behavioral Measures, including self-reported pro-environmental behaviors, intentions, sustainable consumption, and choice preferences; (3) Agent Perception Measures, examining anthropomorphism, animacy, likeability, intelligence, safety, empathy, and climate change impact awareness; and (4) Information Sharing Measures, focusing on willingness to share information and intervention sharing behavior.

By employing regression analyses across these dimensions, we can identify which elements of character-driven narratives most effectively foster connection with marine life and, ultimately, engagement with climate change issues. The robust statistical framework (with R² values ranging from 0.077 to 0.954 across different measures) provides a reliable foundation for understanding the complex relationships between narrative approaches, species selection, and climate change engagement. This structure enables us to trace the pathway from initial perceptions through to actual behavioral outcomes, offering valuable insights for designing effective climate change communication strategies.

\subsubsection{Experimental Conditions and Perceived Variables}
The analysis of perceived anthropomorphism revealed a well-fitted model (R² = 0.0234, p < 0.001). Both Conversational Character Narrative ($\beta = 0.276$, $p < 0.003$) and Static Character Narrative ($\beta = 0.391$ ,$p < 0.001$) conditions demonstrated significantly higher perceived anthropomorphism compared to the Static Scientific Information baseline. This suggests that character-driven narratives, regardless of interactivity, enhanced the perception of human-like qualities in marine life representations.

For perceived animacy, the model showed modest but significant fit ($R² = 0.0111$, $p =0.013$). Notably, the Conversational Character Narrative condition had a negative effect ($\beta = -0.222$, $p =0.010$) compared to the baseline, while the Static Character Narrative showed no significant difference. This unexpected finding suggests that interactive conversations may have highlighted the artificial nature of the experience.

The perceived likeability model demonstrated good fit (R² = 0.0410, p < 0.001). Both Conversational Character Narrative ($\beta = 0.245$, $p =0.007$) and Static Character Narrative ($\beta = 0.476$, $p < 0.001$) conditions significantly improved likeability compared to the baseline, with the static narrative showing a stronger positive effect.

The perceived intelligence analysis showed strong model fit (R² = 0.0615, p < 0.001). Both experimental conditions demonstrated significant positive effects, with Conversational Character Narrative ($\beta = 0.575$, $p < 0.001$) and Static Character Narrative ($\beta = 0.500$, $p < 0.001$) enhancing perceived intelligence of marine life.

For perceived safety, the model was significant (R² = 0.0365, p < 0.001). Both Conversational Character Narrative ($\beta = 0.364$, $p < 0.001$) and Static Character Narrative ($\beta = 0.361$, $p < 0.001$) conditions showed nearly identical positive effects on safety perceptions.

\begin{figure}
    \centering
    \includegraphics[width=1\linewidth]{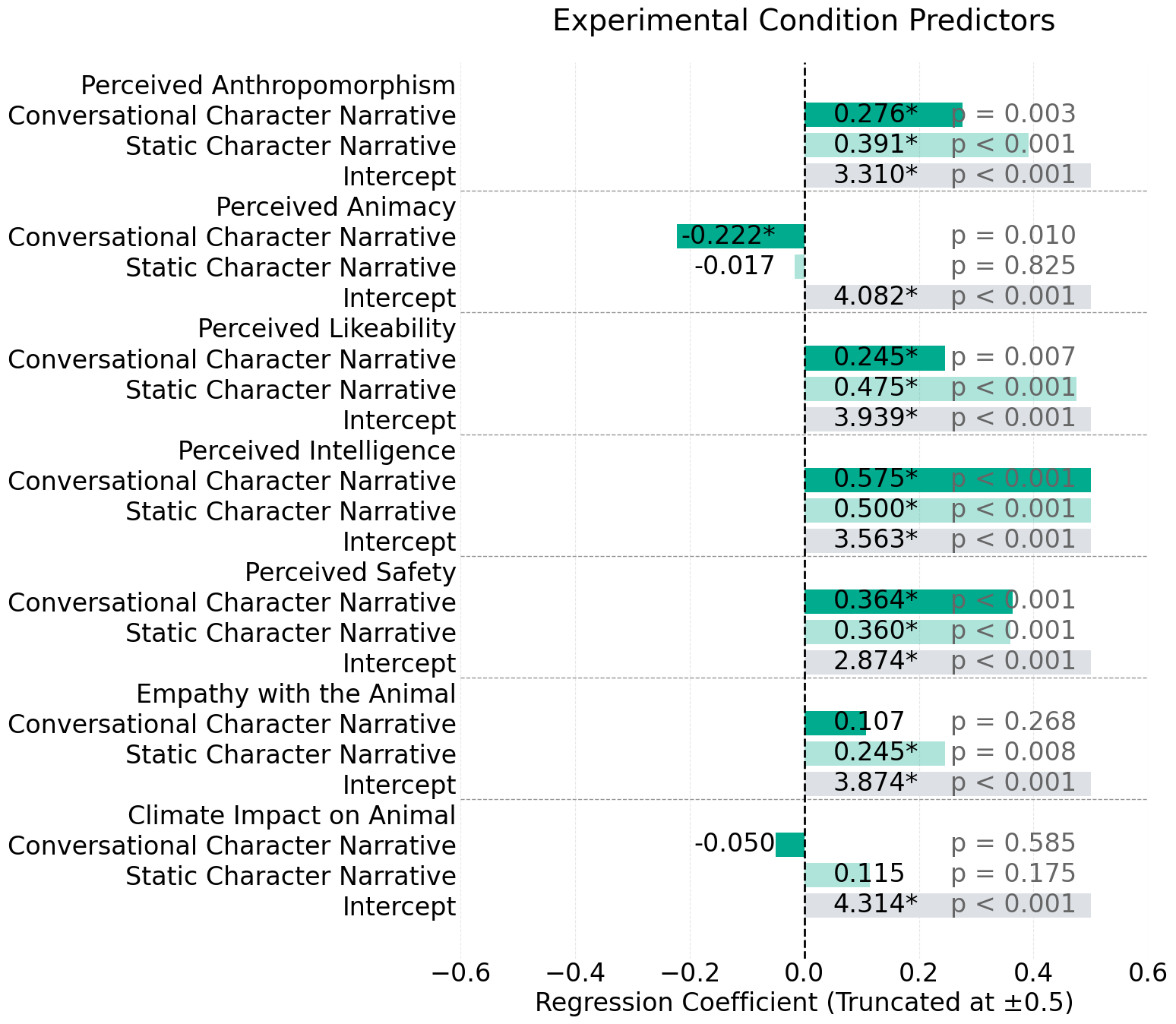}
    \caption{Regression analysis for animal conditions and perceived variables, showing coefficient and p-values}
    \label{fig:animals}
\end{figure}

\subsubsection{Animal Conditions and Perceived Variables}
The analysis of perceived anthropomorphism demonstrates distinct differences between marine species. The beluga and seahorse conditions both showed significant negative effects compared to the jellyfish. The beluga condition had a significant negative effect on perceived anthropomorphism ($\beta = -0.441$, $p < 0.001$), while the seahorse condition also exhibited a significant negative effect ($\beta = -0.456$, $p < 0.001$). These results suggest that marine mammals (e.g., belugas) and small marine vertebrates (e.g., seahorses) are perceived as less anthropomorphic compared to the baseline species (jellyfish).

For perceived animacy, both the beluga ($\beta = -0.172$, $p = 0.037$) and seahorse ($\beta = -0.234$, $p = 0.005$) conditions showed notable differences from the jellyfish baseline. This result contradicts intuitive expectations, as belugas, being highly mobile and socially active marine mammals, are often associated with high levels of animacy due to their dynamic and recognizable behavior. In contrast, jellyfish, typically perceived as slow-moving and less complex in their movements, might be expected to score lower in perceived animacy. The seahorse condition’s lower scores may be partially explained by its relatively static nature and unique mode of movement, which might seem less animated to participants.

The perceived likeability analysis revealed significant positive effects for the beluga condition ($\beta = 0.361$, $p < 0.001$) compared to the jellyfish baseline, with seahorses showing moderate positive effects ($\beta = 0.153$, $p = 0.075$). This pattern suggests that vertebrate species, particularly mammals, may have advantages in fostering positive emotional connections with participants.

Regarding perceived intelligence, the beluga condition showed a strong positive effect ($\beta = 0.280$, $p = 0.001$) compared to the jellyfish baseline, while the seahorse condition showed no significant difference ($\beta = -0.067$, $p = 0.474$). This finding aligns with common perceptions of cetacean cognitive capabilities and may reflect cultural narratives about marine mammal intelligence.

The analysis of perceived safety revealed interesting contrasts, with the beluga condition showing higher perceived safety ($\beta = 0.169$, $p = 0.020$) compared to the jellyfish baseline, while the seahorse condition demonstrated lower perceived safety ($\beta = -0.146$, $p = 0.061$). This pattern may reflect public perceptions of potential threats from different marine species.

The empathy with animals measure showed that the beluga condition elicited significantly higher empathy ($\beta = 0.184$, $p = 0.040$) compared to the jellyfish baseline, while the seahorse condition showed no significant difference ($\beta = -0.061$, $p = 0.544$). This suggests that mammals may have natural advantages in fostering emotional connections with human participants.

For climate change impact awareness, the seahorse condition demonstrated a significantly lower perceived impact awareness ($\beta = -0.233$, $p = 0.007$) compared to the jellyfish baseline, whereas the beluga condition showed no significant difference ($\beta = -0.103$, $p = 0.206$). This finding suggests that certain species may be less effective in conveying vulnerability to climate change, which warrants further investigation.

These findings collectively indicate that while marine mammals may have certain inherent advantages in fostering human connection and engagement, each species presents unique opportunities and challenges for environmental communication. The results highlight the importance of considering species-specific characteristics when designing marine conservation interventions.

\begin{figure}
    \centering
    \includegraphics[width=1\linewidth]{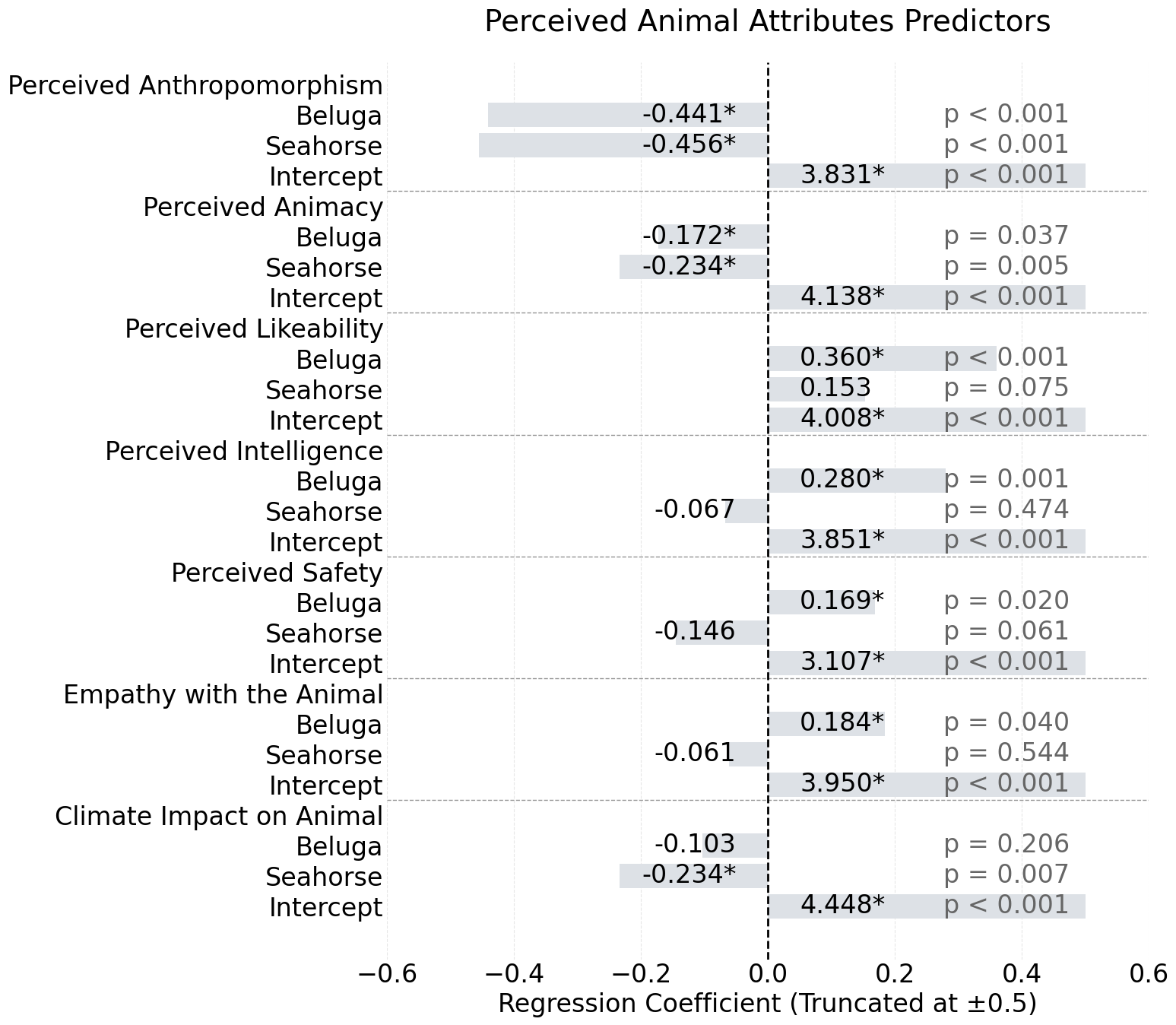}
    \caption{Regression analysis for experimental conditions and perceived variables, showing coefficient and p-values}
    \label{fig:conditions}
\end{figure}

\subsubsection{Climate Change Belief}
The analysis of climate change beliefs revealed a remarkably robust model, with an R-squared value of 0.954 and high statistical significance ($p < 0.001$), indicating that our predictors effectively explained variations in participants' climate change beliefs. The Conversational Character Narrative condition demonstrated a significant positive effect ($\beta = 1.286$, $p = 0.050$) compared to the Static Scientific Information baseline, suggesting that interactive character-driven narratives were more effective at influencing climate change beliefs than traditional scientific communication approaches.

Pre-existing climate change beliefs emerged as the strongest predictor ($\beta = 0.935$, $p < 0.001$), highlighting the importance of accounting for participants' initial perspectives. The analysis revealed several other significant predictors of strengthened climate change beliefs: perceived likeability of the marine characters showed a substantial positive effect ($\beta = 0.783$, $p = 0.026$), while awareness of climate change impacts on marine life demonstrated the strongest positive relationship among the intervention-related variables ($\beta = 2.023$, $p < 0.001$).

Gender showed a marginally significant positive relationship ($\beta = 0.771$, $p = 0.082$), suggesting potential demographic variations in receptiveness to the intervention. Notably, neither education level nor location demonstrated significant effects on climate change beliefs, indicating that the intervention's impact transcended these demographic factors.

When examining the animal conditions, neither the jellyfish ($\beta = 0.673$, $p = 0.244$) nor seahorse ($\beta = 0.783$, $p = 0.157$) conditions showed significant differences compared to the baseline, suggesting that the species of marine life featured did not significantly influence the intervention's effect on climate change beliefs. This finding indicates that the narrative approach and interaction style may be more crucial than the specific marine species represented in shaping climate change beliefs.

These results demonstrate that interactive, character-driven narratives can effectively influence climate change beliefs, particularly when they successfully establish emotional connections through character likeability and effectively communicate climate change impacts on marine ecosystems. The high model fit suggests that this approach offers a promising direction for climate change communication and education efforts.

\begin{figure}
    \centering
    \includegraphics[width=1\linewidth]{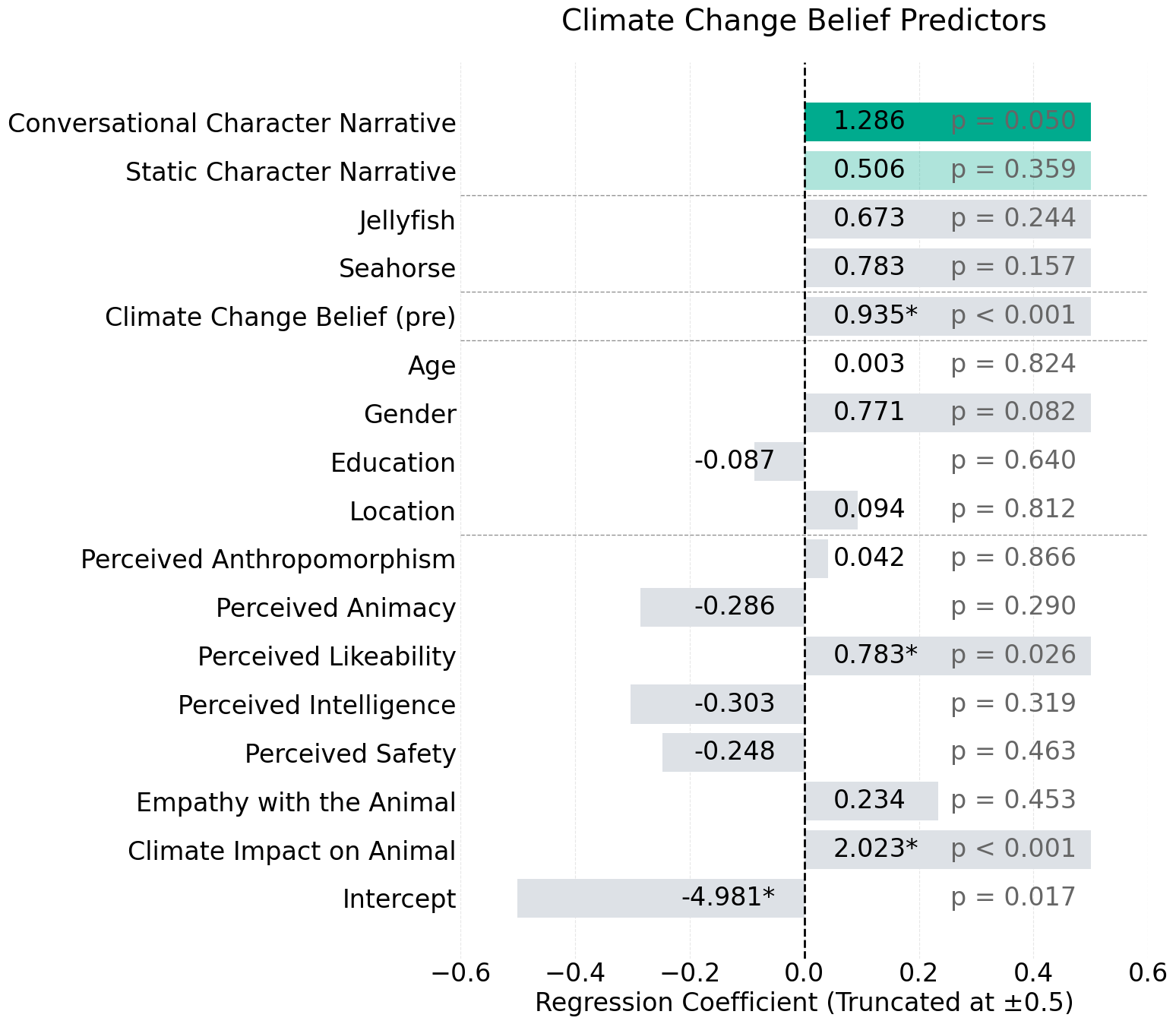}
    \caption{Regression analysis for experimental conditions and climate change belief, showing coefficient and p-values}
    \label{fig:belief_post}
\end{figure}

\subsubsection{Psychological Distance Toward Climate Change}

The analysis of psychological distance toward climate change revealed a robust model fit ($R^2 = 0.678$, $\text{Adj } R^2 = 0.670$, $p < 0.001$), indicating that our predictors effectively captured variations in how participants perceived their personal connection to climate change issues.

Pre-existing psychological distance emerged as the strongest predictor in the model ($\beta = 0.861$, $p < 0.001$), demonstrating the substantial influence of participants' baseline perspectives. However, neither the Conversational Character Narrative ($\beta = 0.020$, $p = 0.717$) nor the Static Character Narrative ($\beta = 0.064$, $p = 0.164$) conditions showed significant differences compared to the Static Scientific Information baseline, suggesting that the narrative approach alone may not be sufficient to shift psychological distance.

Geographic location demonstrated a significant positive relationship with psychological distance ($\beta = 0.066$, $p = 0.021$), indicating that spatial factors play a meaningful role in how individuals relate to climate change issues. The analysis also revealed that awareness of climate change impacts on marine life had a significant positive effect ($\beta = 0.097$, $p = 0.001$), suggesting that understanding specific environmental consequences may help bridge the psychological gap to climate change.

Regarding the animal conditions, neither the jellyfish ($\beta = 0.055$, $p = 0.253$) nor seahorse ($\beta = 0.006$, $p = 0.899$) conditions showed significant differences in affecting psychological distance compared to the baseline. This suggests that the species of marine life featured did not significantly influence participants' sense of connection to climate change issues.

Notably, demographic factors such as age ($\beta = 0.001$, $p = 0.819$), gender ($\beta = 0.034$, $p = 0.347$), and education ($\beta = 0.007$, $p = 0.607$) did not demonstrate significant effects on psychological distance. Similarly, perceived characteristics of the marine life characters, including anthropomorphism, animacy, likeability, intelligence, and safety, showed no significant influence on psychological distance.

These findings suggest that while the intervention successfully engaged participants, reducing psychological distance toward climate change may require additional strategies beyond character-driven narratives. The significant effects of location and climate change impact awareness indicate that focusing on local relevance and concrete environmental consequences may be particularly effective approaches for future interventions.

\begin{figure}
    \centering
    \includegraphics[width=1\linewidth]{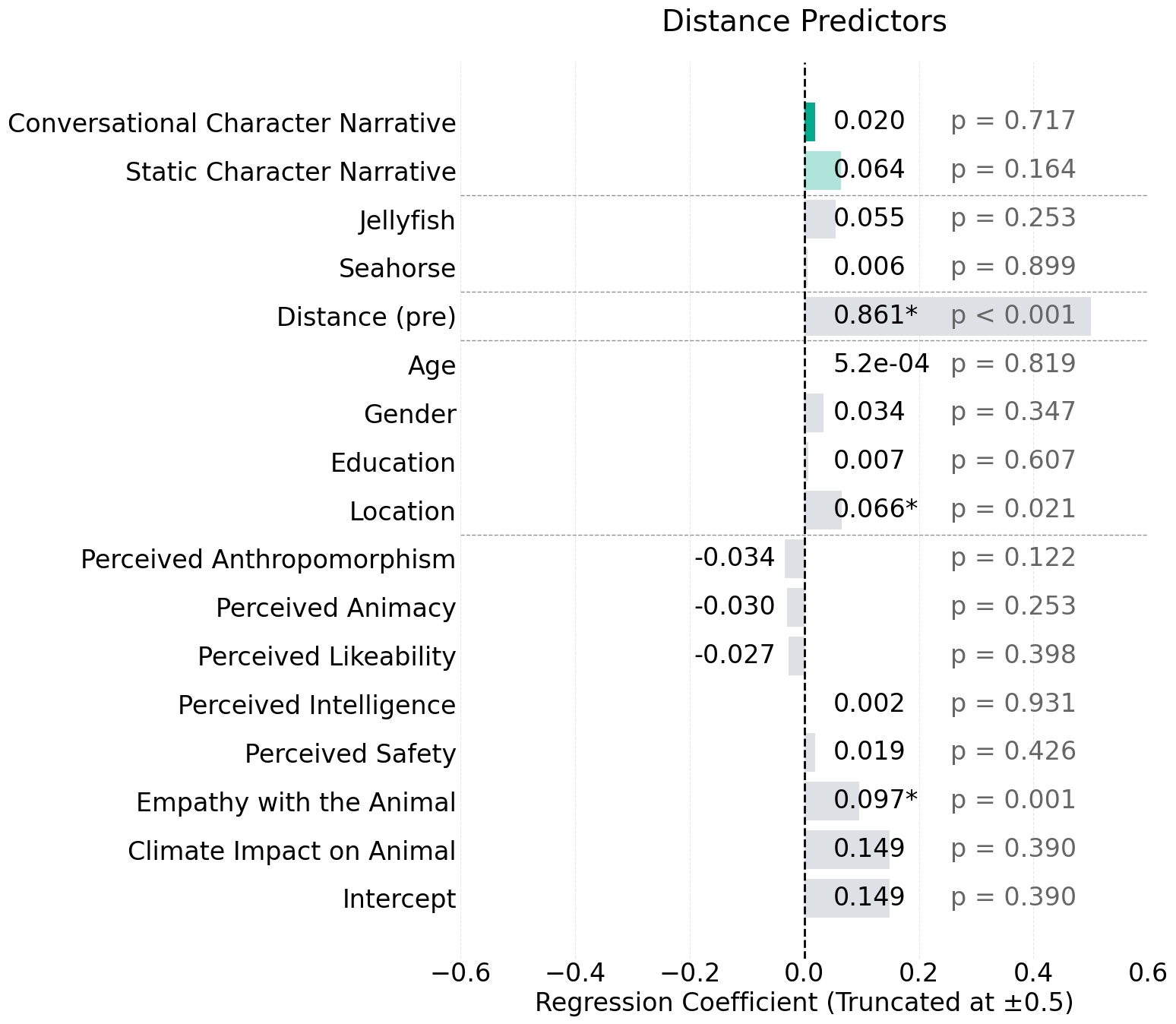}
    \caption{Regression analysis for experimental conditions and psychological distance, showing coefficient and p-values}
    \label{fig:distance_post}
\end{figure}

\subsubsection{Climate Policy Adoption}

The analysis of climate policy adoption attitudes revealed an exceptionally strong model fit ($R^2 = 0.947$, $\text{Adj } R^2 = 0.946$, $p < 0.001$), indicating that our predictors effectively explained variations in participants' support for climate-related policies.

Pre-existing climate policy attitudes emerged as the dominant predictor ($\beta = 0.961$, $p < 0.001$), demonstrating the strong influence of participants' initial policy positions. Notably, awareness of climate change impacts on marine life showed a substantial positive effect ($\beta = 0.940$, $p = 0.002$), suggesting that understanding specific environmental consequences significantly influenced policy support.

Neither the Conversational Character Narrative ($\beta = 0.315$, $p = 0.535$) nor the Static Character Narrative ($\beta = -0.379$, $p = 0.394$) conditions demonstrated significant differences compared to the Static Scientific Information baseline. This indicates that the narrative approach alone may not be sufficient to shift established policy preferences.

Location showed a marginally positive relationship with policy support ($\beta = 0.488$, $p = 0.117$), suggesting that geographic factors may influence climate policy attitudes. Empathy with marine animals also demonstrated a notable positive relationship ($\beta = 0.441$, $p = 0.131$), although not reaching statistical significance.

Regarding the animal conditions, neither the jellyfish ($\beta = 0.066$, $p = 0.888$) nor seahorse ($\beta = 0.379$, $p = 0.402$) conditions significantly influenced policy support compared to the baseline. This suggests that the specific marine species featured did not substantially affect participants' policy preferences.

Demographic factors including age ($\beta = 0.009$, $p = 0.715$), gender ($\beta = 0.096$, $p = 0.792$), and education ($\beta = -0.056$, $p = 0.723$) showed no significant effects on climate policy support. Similarly, perceived characteristics of the marine life characters (anthropomorphism, animacy, likeability, intelligence, and safety) did not significantly influence policy attitudes.

These findings suggest that while the intervention successfully engaged participants, shifting climate policy preferences may require approaches that go beyond character-driven narratives. The strong influence of pre-existing attitudes and the significant effect of understanding climate change impacts indicate that future interventions might be more effective by focusing on concrete environmental consequences and their policy implications.

\begin{figure}
    \centering
    \includegraphics[width=1\linewidth]{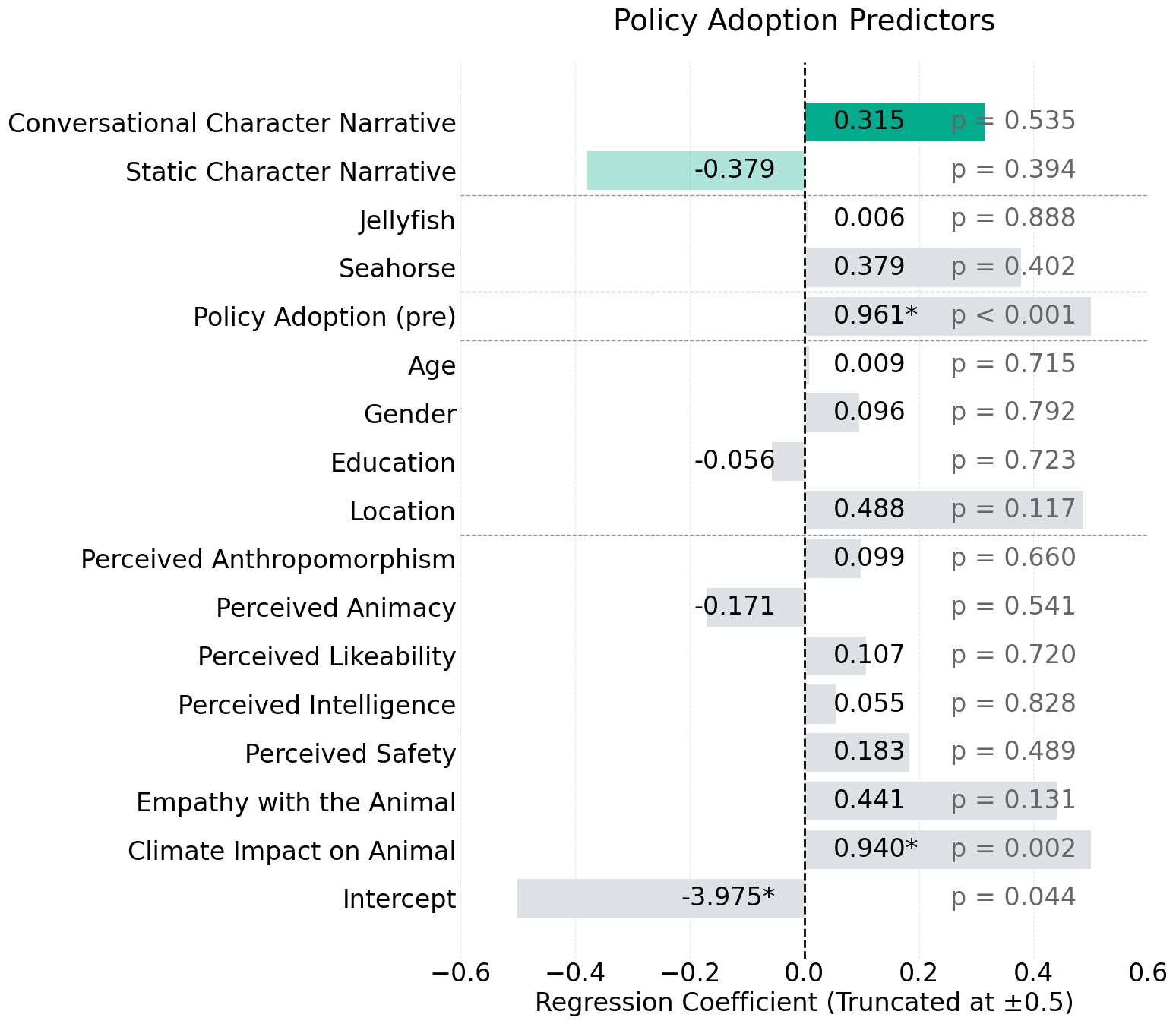}
    \caption{Regression analysis for experimental conditions and policy adoption, showing coefficient and p-values}
    \label{fig:policy_post}
\end{figure}

\subsubsection{Self-Reported Pro-Environmental Behaviors}

The analysis of self-reported pro-environmental behavior demonstrated an excellent model fit (\(R^2 = 0.9764\), \(\text{Adj } R^2 = 0.9759\), \(p < 0.001\)), indicating that our predictors explained the majority of the variance in participants' reported behaviors.

Pre-existing pro-environmental behavior was the strongest predictor (\(\beta = 0.996\), \(p < 0.001\)), underscoring the high stability of participants' prior behavior in predicting post-intervention behavior. Neither the Conversational Character Narrative (\(\beta = 0.018\), \(p = 0.741\)) nor the Static Character Narrative (\(\beta = -0.006\), \(p = 0.197\)) significantly affected self-reported behavior compared to the baseline condition.

No significant effects were observed for the animal conditions. Both the jellyfish (\(\beta = -0.005\), \(p = 0.913\)) and seahorse (\(\beta = 0.015\), \(p = 0.144\)) conditions failed to yield significant differences in self-reported behavior compared to the baseline.

Demographic predictors, including age (\(\beta = 0.0003\), \(p = 0.784\)), gender (\(\beta = 0.012\), \(p = 0.689\)), education (\(\beta = -0.003\), \(p = 0.752\)), and location (\(\beta = 0.008\), \(p = 0.554\)), also showed no significant influence on self-reported behavior.

Among perception-related predictors, no significant effects were detected for perceived anthropomorphism (\(\beta = -0.0008\), \(p = 0.913\)), perceived animacy (\(\beta = 0.006\), \(p = 0.817\)), perceived likeability (\(\beta = 0.009\), \(p = 0.426\)), or perceived intelligence (\(\beta = -0.007\), \(p = 0.217\)). Empathy with marine animals (\(\beta = 0.011\), \(p = 0.243\)) and perceived safety (\(\beta = -0.005\), \(p = 0.689\)) were also not significant predictors.

Awareness of climate change impacts on marine life showed a marginally significant positive effect (\(\beta = 0.060\), \(p = 0.064\)), suggesting a potential trend for this variable to influence self-reported pro-environmental behavior.

These findings align with expectations, as participants were unlikely to change their past behavioral perceptions significantly within the short pre-post intervention period. Such consistency in self-reported behavior is expected because altering previous actions in such a brief time would create cognitive dissonance. Participants’ responses reflect the stability of behavioral patterns, reaffirming the challenge of achieving immediate behavioral change through interventions targeting pro-environmental actions.
\begin{figure}
    \centering
    \includegraphics[width=1\linewidth]{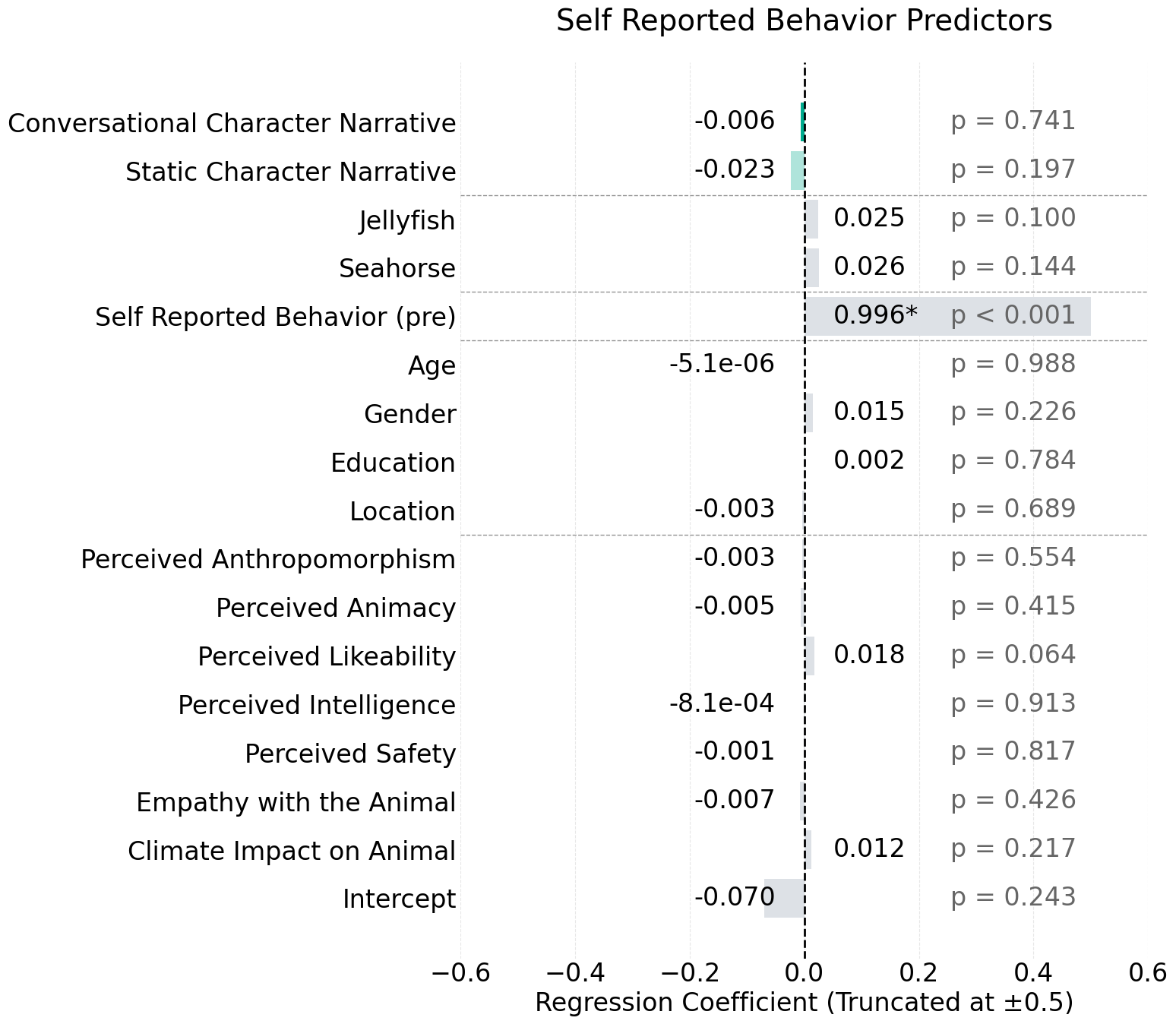}
    \caption{Regression analysis for experimental conditions and self-reported pro-environmental intentions, showing coefficient and p-values}
    \label{fig:behavior_post}
\end{figure}

\subsubsection{Pro-Environmental Intentions}

The analysis of pro-environmental behaviors demonstrated a robust model fit ($R^2 = 0.792$, $\text{Adj } R^2 = 0.788$, $p < 0.001$), indicating that our predictors effectively explained variations in participants' self-reported environmental actions.

The Conversational Character Narrative condition showed a highly significant positive effect ($\beta = 0.173$, $p < 0.001$) compared to the Static Scientific Information baseline. The Static Character Narrative condition showed a positive but non-significant effect ($\beta = 0.039$, $p = 0.315$), further suggesting that interactivity may influence how participants set pro-environmental intentions.

Pre-existing pro-environmental intentions emerged as the strongest predictor ($\beta = 0.875$, $p < 0.001$), emphasizing the importance of baseline environmental attitudes in shaping self-reported behaviors. Several perception-related variables showed significant positive relationships with reported pro-environmental behaviors: perceived animacy ($\beta = 0.036$, $p = 0.069$), perceived likeability ($\beta = 0.059$, $p = 0.012$), and awareness of climate change impacts on marine life ($\beta = 0.044$, $p = 0.009$). These results suggest that participants' perceptions of the intervention may have influenced their reporting rather than reflecting objective behavioral changes.

Regarding the animal conditions, neither the jellyfish ($\beta = -0.003$, $p = 0.952$) nor seahorse ($\beta = 0.011$, $p = 0.781$) conditions significantly influenced self-reported pro-environmental behaviors compared to the baseline. This suggests that the narrative approach and interaction style were more crucial than the specific marine species featured in shaping participants' responses.

Demographic factors including age ($\beta = 0.003$, $p = 0.831$), gender ($\beta = 0.022$, $p = 0.531$), education ($\beta = -0.018$, $p = 0.175$), and location ($\beta = 0.012$, $p = 0.663$) showed no significant effects on self-reported behaviors. Similarly, perceived intelligence ($\beta = -0.003$, $p = 0.887$) and perceived safety ($\beta = 0.003$, $p = 0.895$) did not significantly influence reported behavioral outcomes.

Despite the significance of pre-existing intentions the conversational chat interface could significantly increase pro-environmental intentions. With likeability and climate change impact on the animal as further predictors this confirms the assumption that conversational narratives, depicted by animals have the potential to significantly facilitate PEB.

\begin{figure}
    \centering
    \includegraphics[width=1\linewidth]{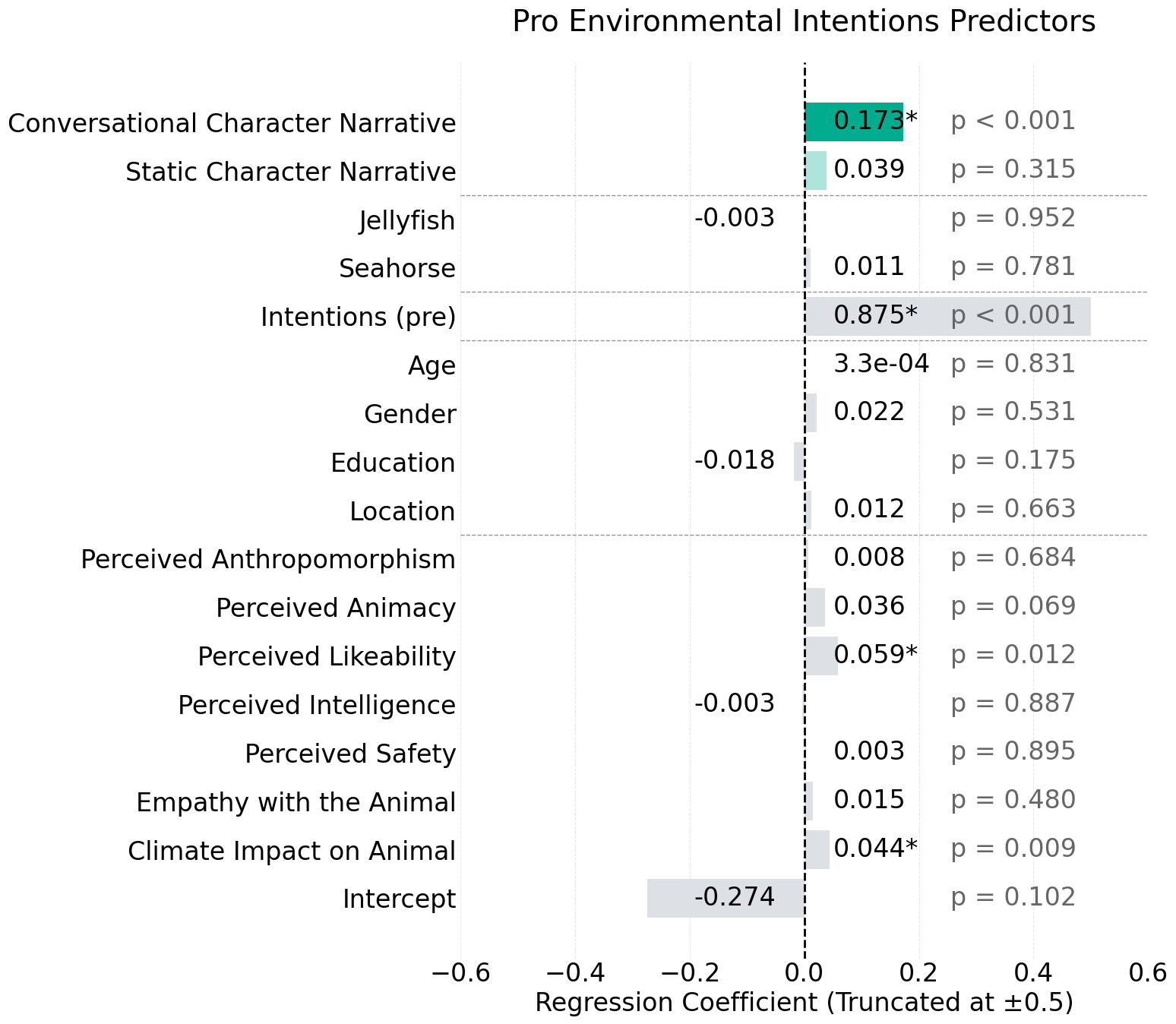}
    \caption{Regression analysis for experimental conditions and pro-environmental intentions, showing coefficient and p-values}
    \label{fig:intentions_post}
\end{figure}

\subsubsection{Sustainable Consumption}
The analysis of sustainable consumption patterns demonstrated a remarkably strong model fit ($R^2 = 0.904$, $\text{Adj } R^2 = 0.902$, $p < 0.001$), indicating that our predictors effectively captured variations in participants' sustainable consumption behaviors.

The Conversational Character Narrative condition showed a marginally significant positive effect ($\beta = 0.077$, $p = 0.079$) compared to the Static Scientific Information baseline, suggesting that interactive character-driven narratives may have some advantage in promoting sustainable consumption. In contrast, the Static Character Narrative condition showed no significant effect ($\beta = -0.022$, $p = 0.578$).

Pre-existing sustainable consumption patterns emerged as the strongest predictor ($\beta = 0.926$, $p < 0.001$), demonstrating the substantial influence of established behaviors. Notably, several intervention-related factors showed significant positive relationships with sustainable consumption: perceived anthropomorphism ($\beta = 0.036$, $p = 0.041$), perceived intelligence ($\beta = 0.043$, $p = 0.082$, marginally significant), and awareness of climate change impacts on marine life ($\beta = 0.069$, $p = 0.007$).

Location showed a marginally significant positive relationship with sustainable consumption ($\beta = 0.042$, $p = 0.113$), suggesting that geographic factors may play a role in consumption patterns. Similarly, empathy with marine animals demonstrated a positive relationship ($\beta = 0.039$, $p = 0.113$), though not reaching conventional levels of statistical significance.

Regarding the animal conditions, neither the jellyfish ($\beta = 0.055$, $p = 0.143$) nor seahorse ($\beta = 0.054$, $p = 0.161$) conditions significantly influenced sustainable consumption compared to the baseline. This suggests that the specific marine species featured did not substantially affect participants' consumption choices.

Demographic factors including age ($\beta = 0.001$, $p = 0.769$), gender ($\beta = 0.031$, $p = 0.318$), and education ($\beta = 0.001$, $p = 0.966$) showed no significant effects on sustainable consumption patterns. Perceived safety ($\beta = -0.026$, $p = 0.231$) and perceived animacy ($\beta = -0.015$, $p = 0.434$) also did not significantly influence consumption behaviors.

These findings indicate that while interactive character narratives show promise in promoting sustainable consumption, their effectiveness may be enhanced when they successfully generate anthropomorphic connections and effectively communicate climate change impacts. The strong influence of pre-existing consumption patterns suggests that interventions might be most effective when designed to reinforce and extend existing sustainable practices rather than attempting to create entirely new behavior patterns.
\begin{figure}
    \centering
    \includegraphics[width=1\linewidth]{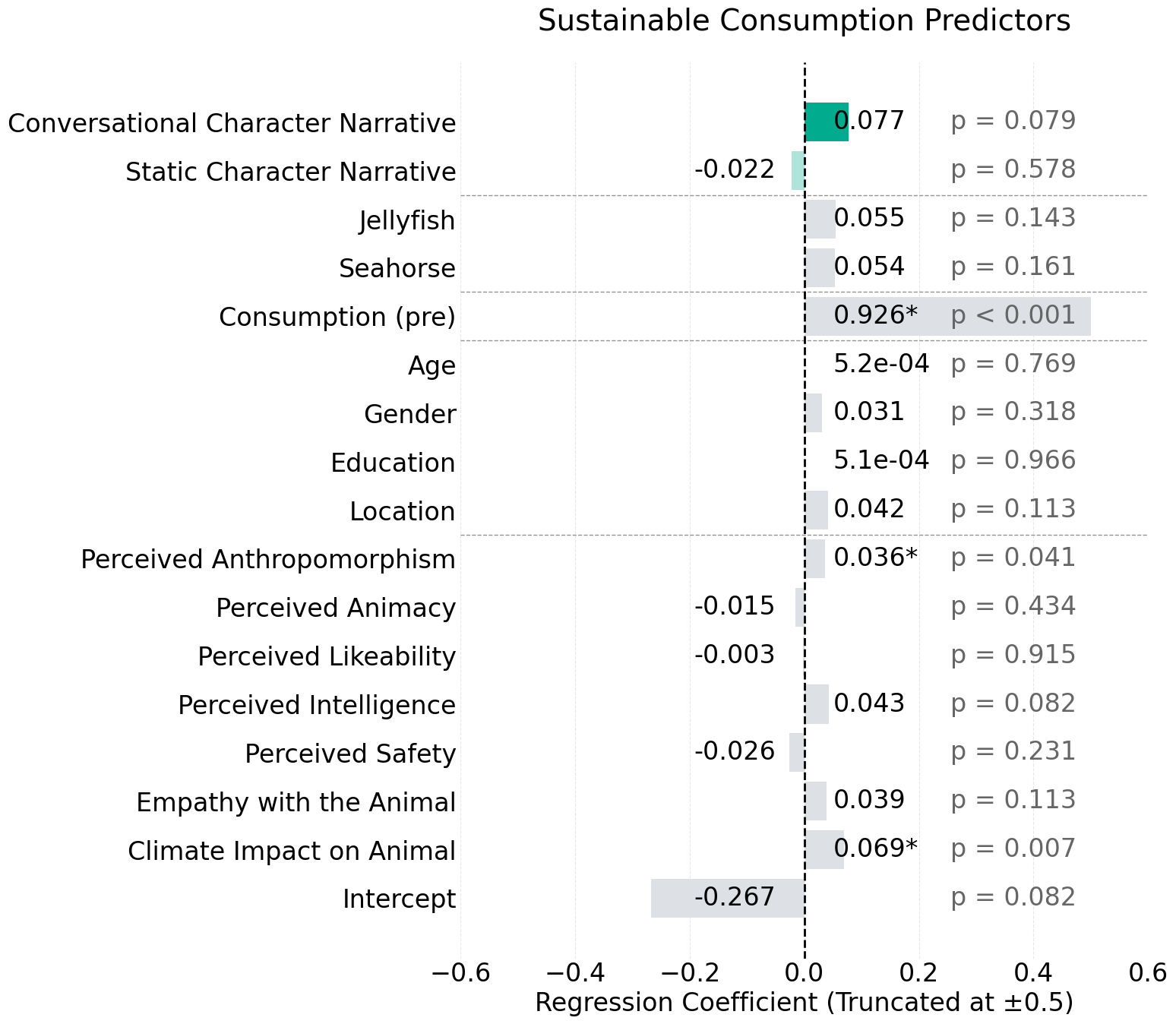}
    \caption{Regression analysis for experimental conditions and sustainable consumption, showing coefficient and p-values}
    \label{fig:consumption_post}
\end{figure}

\subsubsection{Sustainable Choice Preferences}
The analysis of sustainable choice preferences revealed a modest model fit ($R^2 = 0.192$, $\text{Adj } R^2 = 0.175$, $p < 0.001$), indicating that the predictors explained a limited but meaningful portion of the variance in participants' preferences for sustainable choices.

The Conversational Character Narrative condition demonstrated a significant positive effect ($\beta = 0.342$, $p = 0.031$) compared to the Static Scientific Information baseline, suggesting that interactive, character-driven narratives positively influenced participants' sustainable consumption patterns. Conversely, the Static Character Narrative condition showed no significant effect ($\beta = 0.115$, $p = 0.434$), indicating that interactivity may play a crucial role in shaping preferences for sustainable options.

Several perception-related variables emerged as significant predictors. Perceived animacy ($\beta = 0.174$, $p = 0.007$) and perceived likeability ($\beta = 0.142$, $p = 0.075$) both demonstrated positive relationships with sustainable choice preferences. Additionally, empathy with marine animals showed a strong and highly significant positive effect ($\beta = 0.404$, $p < 0.001$), emphasizing the role of emotional connection in driving participants’ preferences for sustainability.

In contrast, perceived anthropomorphism ($\beta = 0.025$, $p = 0.905$), perceived intelligence ($\beta = -0.035$, $p = 0.663$), and perceived safety ($\beta = -0.057$, $p = 0.451$) did not significantly influence sustainable choice preferences. These findings suggest that while certain perception-related variables are impactful, others may have less relevance in shaping sustainable behaviors.

Regarding the animal conditions, neither the jellyfish ($\beta = 0.085$, $p = 0.596$) nor the seahorse ($\beta = -0.025$, $p = 0.862$) conditions significantly influenced participants' preferences compared to the beluga, further suggesting that the narrative approach is more critical than the specific species featured.

Demographic variables showed limited influence on sustainable choice preferences. Age ($\beta = 0.0004$, $p = 0.905$) and gender ($\beta = 0.182$, $p = 0.132$) were not significant predictors, while education ($\beta = -0.059$, $p = 0.226$) and location ($\beta = 0.224$, $p = 0.029$) showed marginal effects, with location reaching significance. These results indicate that geographic factors may moderately influence sustainable choice preferences, though demographic variables overall played a limited role.

\begin{figure}
    \centering
    \includegraphics[width=1\linewidth]{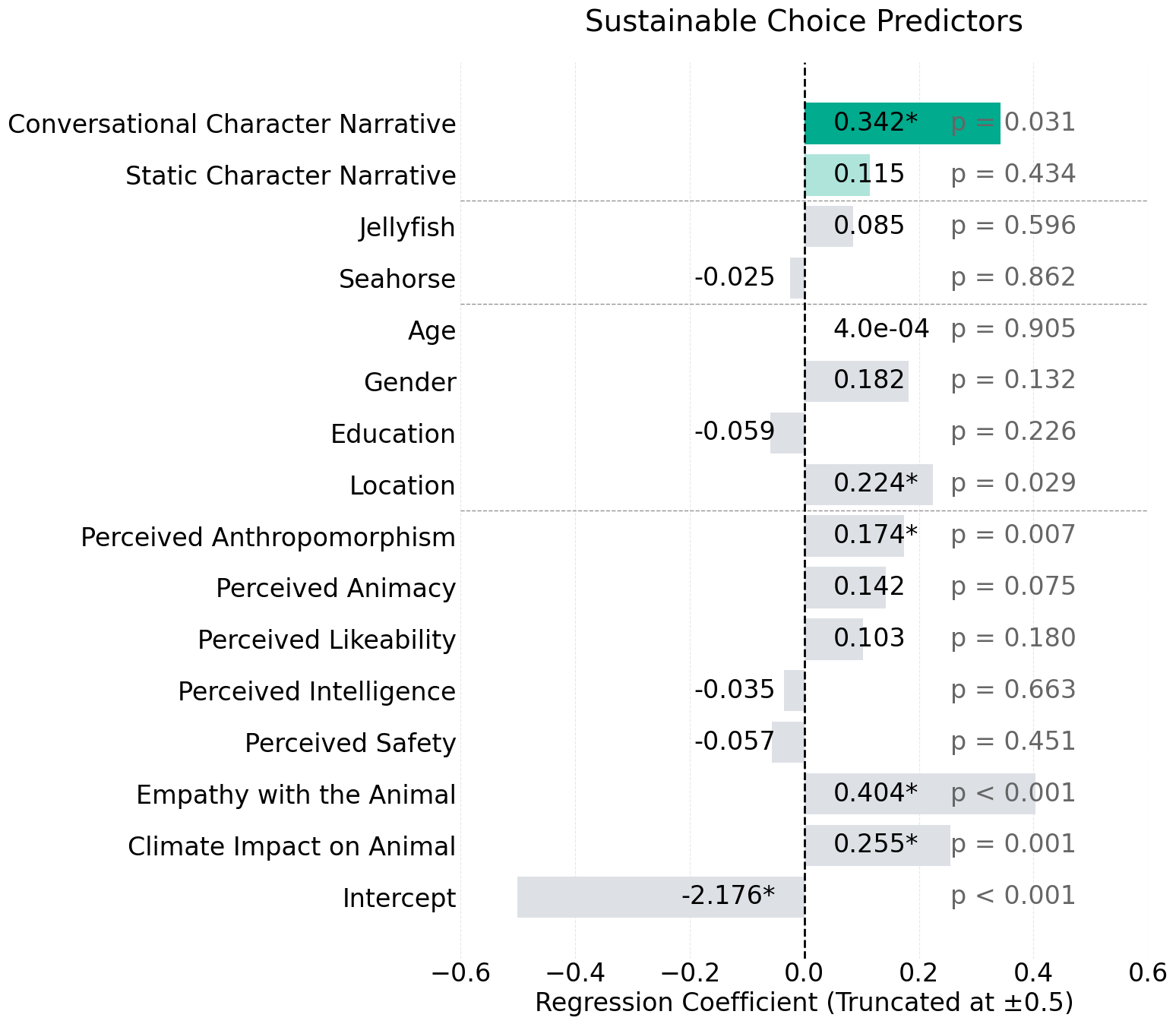}
    \caption{Regression analysis for experimental conditions and sustainable choice preferences, showing coefficient and p-values}
    \label{fig:choices_post}
\end{figure}
\subsubsection{Willingness to Share Climate Change Information}

The analysis of willingness to share climate change information revealed a strong model fit ($R^2 = 0.811$, $\text{Adj } R^2 = 0.807$, $p < 0.001$), indicating robust predictive power for understanding participants' information sharing tendencies.

Pre-existing willingness to share emerged as the dominant predictor ($\beta = 0.880$, $p < 0.001$), demonstrating the strong influence of participants' baseline sharing preferences. Neither the Conversational Character Narrative ($\beta = 0.010$, $p = 0.626$) nor the Static Character Narrative ($\beta = 0.004$, $p = 0.830$) conditions showed significant differences compared to the Static Scientific Information baseline, suggesting that narrative approach alone did not substantially influence sharing behavior.

Two perception-related variables demonstrated significant positive effects on sharing willingness: perceived animacy ($\beta = 0.025$, $p = 0.017$) and awareness of climate change impacts on marine life ($\beta = 0.023$, $p = 0.005$). Additionally, gender showed a marginally significant positive relationship ($\beta = 0.026$, $p = 0.095$), suggesting potential demographic variations in sharing behavior.

Perceived anthropomorphism showed a marginally significant negative relationship ($\beta = -0.020$, $p = 0.072$), an unexpected finding that warrants further investigation. Regarding the animal conditions, neither the jellyfish ($\beta = 0.008$, $p = 0.671$) nor seahorse ($\beta = 0.016$, $p = 0.474$) conditions significantly influenced sharing willingness compared to the baseline.

Demographic factors including age ($\beta = -0.0003$, $p = 0.933$), education ($\beta = -0.009$, $p = 0.125$), and location ($\beta = 0.008$, $p = 0.545$) showed no significant effects on sharing behavior. Other perceived characteristics such as intelligence ($\beta = -0.004$, $p = 0.768$), safety ($\beta = -0.012$, $p = 0.234$), and empathy with animals ($\beta = 0.007$, $p = 0.520$) also did not significantly influence sharing willingness.

\begin{figure}
    \centering
    \includegraphics[width=1\linewidth]{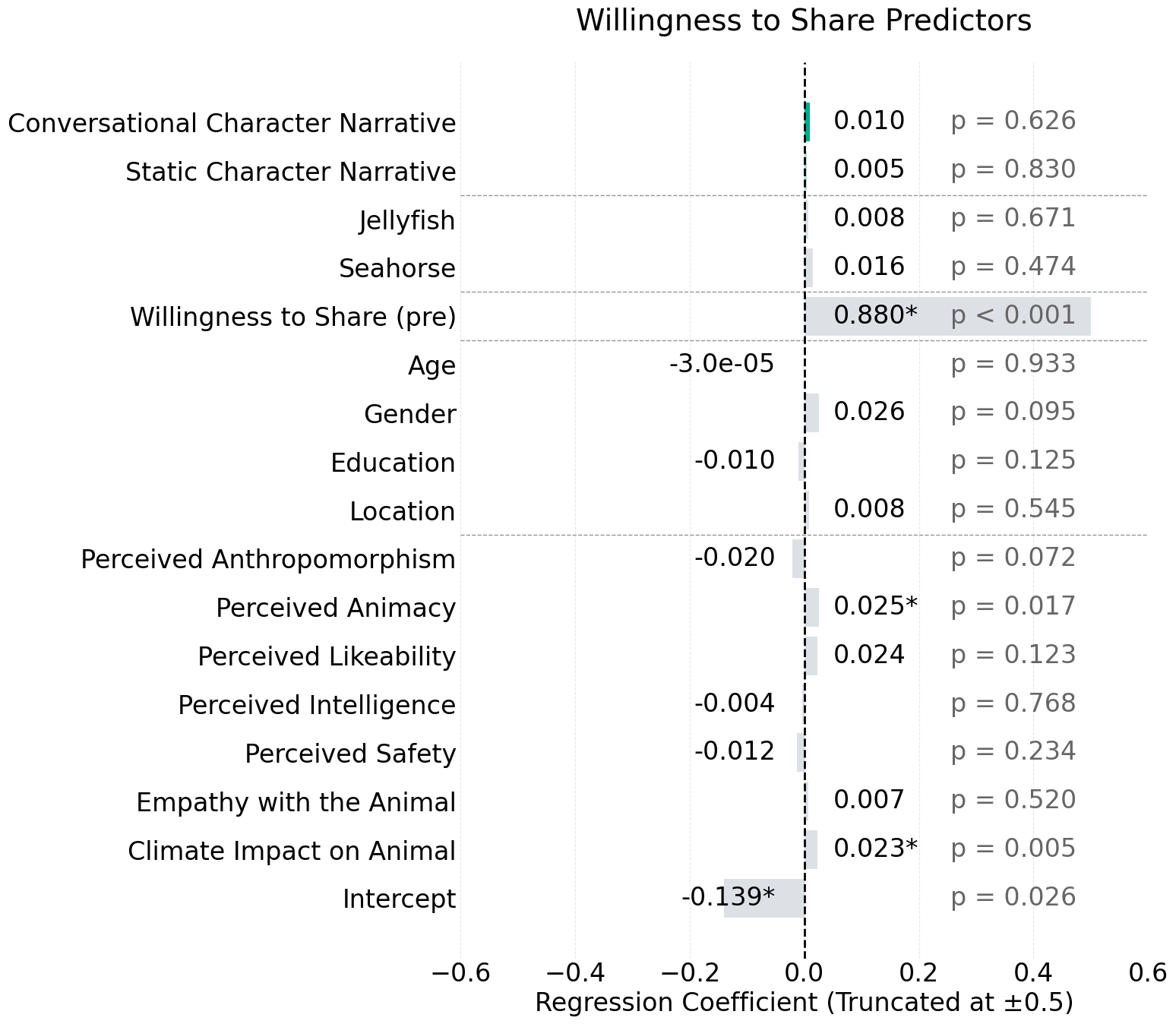}
    \caption{Regression analysis for experimental conditions and willingness to share climate change information, showing coefficient and p-values}
    \label{fig:sharing_post}
\end{figure}

\subsubsection{Intervention Sharing}

The analysis of intervention sharing behavior revealed a modest but significant model fit ($R^2 = 0.077$, $\text{Adj } R^2 = 0.047$, $p < 0.001$), indicating that our predictors explained a small but meaningful portion of variance in participants' willingness to share the intervention with others.

The experimental conditions showed no significant effects, with neither the Conversational Character Narrative ($\beta = 0.059$, $p = 0.324$) nor Static Character Narrative ($\beta = 0.002$, $p = 0.977$) conditions differing significantly from the Static Scientific Information baseline. This suggests that the narrative approach alone did not substantially influence sharing behavior.

Several perception-related variables emerged as significant predictors. Empathy with the marine animals demonstrated a significant positive effect ($\beta = 0.065$, $p = 0.041$), indicating that emotional connection played an important role in sharing behavior. Perceived animacy also showed a marginally significant positive relationship ($\beta = 0.049$, $p = 0.098$), suggesting that the perception of marine life as autonomous beings influenced sharing tendencies.

Perceived intelligence showed a positive but non-significant relationship ($\beta = 0.048$, $p = 0.148$), while perceived anthropomorphism ($\beta = 0.025$, $p = 0.318$) and perceived likeability ($\beta = 0.008$, $p = 0.827$) had minimal effects on sharing behavior. Notably, awareness of climate change impacts on marine life showed a negative but non-significant relationship ($\beta = -0.183$, $p = 0.563$).

Regarding the animal conditions, neither the jellyfish ($\beta = -0.007$, $p = 0.896$) nor seahorse ($\beta = -0.065$, $p = 0.219$) conditions significantly influenced sharing behavior compared to the baseline. Demographic factors including age ($\beta = -0.0001$, $p = 0.906$), gender ($\beta = -0.034$, $p = 0.429$), education ($\beta = 0.019$, $p = 0.265$), and location ($\beta = 0.030$, $p = 0.403$) showed no significant effects.

These findings suggest that while the intervention generated engagement, the willingness to share it was primarily driven by emotional connection and perceived animacy rather than narrative approach or specific marine species featured. Future interventions aiming to promote sharing behavior might benefit from focusing on fostering emotional connections and highlighting the autonomous nature of marine life.

\begin{figure}
    \centering
    \includegraphics[width=1\linewidth]{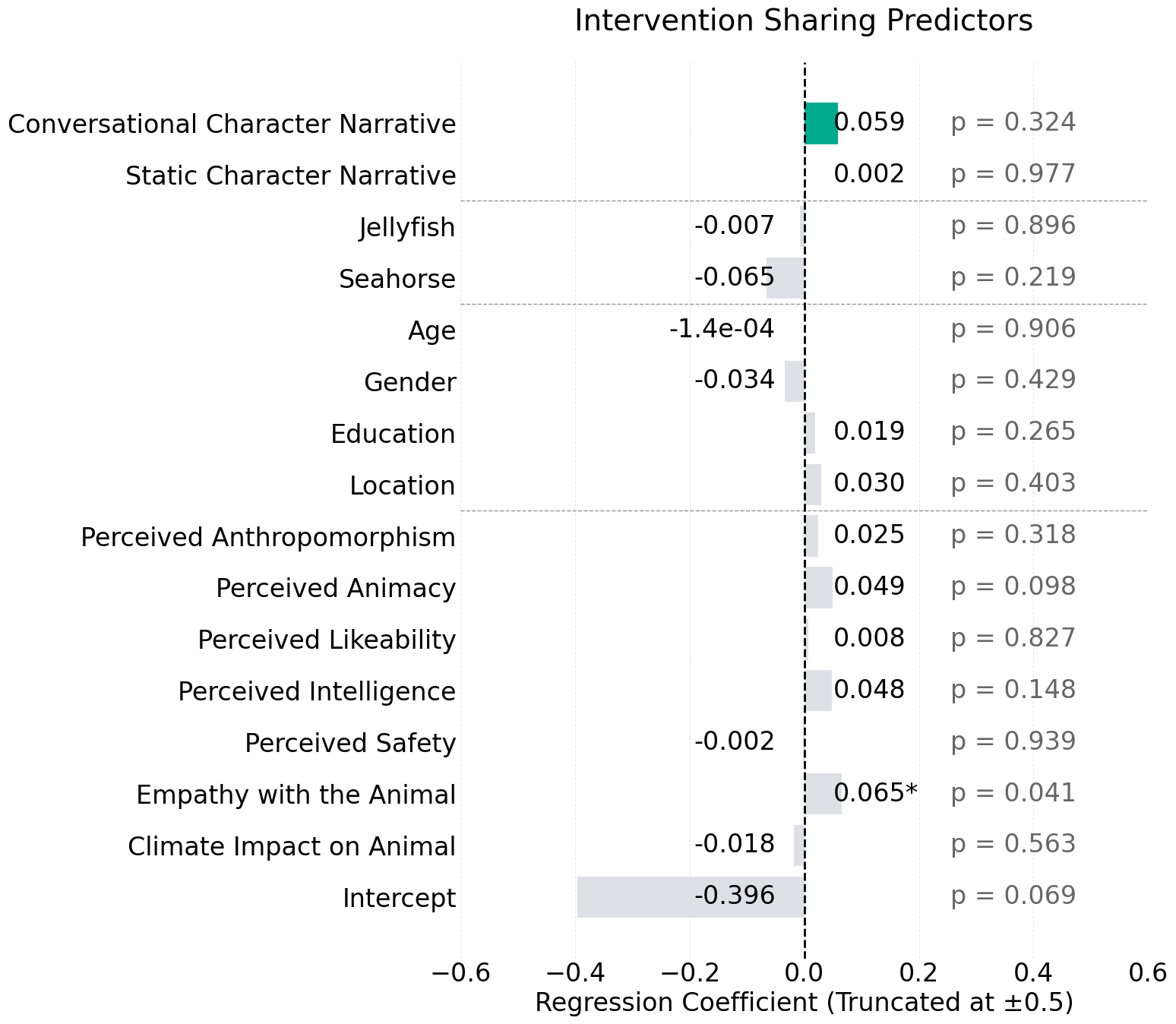}
    \caption{Regression analysis for experimental conditions and intervention sharing, showing coefficient and p-values}
    \label{fig:intervention_sharing_post}
\end{figure}

\subsection{Qualitative Analysis}
The qualitative analysis was conducted based on the chat responses of OceanChat, which were anonymously logged for each participant with the respective condition. This qualitative analysis examines participant responses from OceanChat interactions, revealing several key themes in environmental attitudes and behaviors. The study identified five primary areas: pro-environmental behavioral intentions, emotional resonance and empathy towards environmental issues, skepticism about individual impact, socioeconomic constraints affecting sustainable choices, and the presence of cognitive dissonance in environmental behaviors. While many participants expressed strong intentions to adopt sustainable practices and demonstrated an emotional connection to environmental causes, they also acknowledged significant barriers to implementation, including financial limitations and systemic constraints. The analysis highlights a notable tension between individual responsibility and the perceived need for broader institutional change, with some participants questioning the efficacy of personal actions in addressing environmental challenges.

\subsubsection{Pro-Environmental Behavioral Intentions}
In the responses gathered, a diverse range of pro-environmental behavioral intentions emerged, reflecting both individual and community-level commitments to sustainability. Many respondents expressed clear intentions to reduce their reliance on single-use plastics—one participant noted, “Yes, I'll skip the straw next time,” while several others mentioned already using reusable water bottles, metal straws, or reusable bags. Beyond these direct actions, numerous individuals emphasized broader strategies such as recycling rigorously, opting for refillable alternatives in everyday purchases, and even modifying dietary habits (e.g., eating less meat and dairy) in order to lower their carbon footprints. A number of participants underscored the importance of spreading environmental awareness through social media and community engagement, suggesting that personal behavior changes could serve as a catalyst for broader societal shifts. Some responses acknowledged the limitations of individual efforts alone, with remarks noting that systemic change—through corporate accountability and government intervention—is necessary to address issues like plastic pollution and climate change. Nonetheless, the overarching sentiment was one of proactive commitment: respondents demonstrated a willingness to adjust daily routines, from carpooling and reducing vehicle use to opting for bulk purchases over heavily packaged goods, with the goal of mitigating environmental degradation and promoting a more sustainable future.

\subsubsection{Emotional Resonance and Empathy}
Participants frequently conveyed a profound sense of emotional resonance and empathy in their responses, deeply connecting with the environmental issues discussed. Many exhibited heartfelt concern for the well-being of nature, often expressing sorrow and regret over the damage inflicted upon marine life and ecosystems. This emotional response was reflected in the findings of the sentiment analysis conducted using the \textit{text2emotion} package \cite{Gupta2021}, which highlighted sadness as the most dominant emotion, followed by fear.

\begin{figure}
    \centering
    \includegraphics[width=1\linewidth]{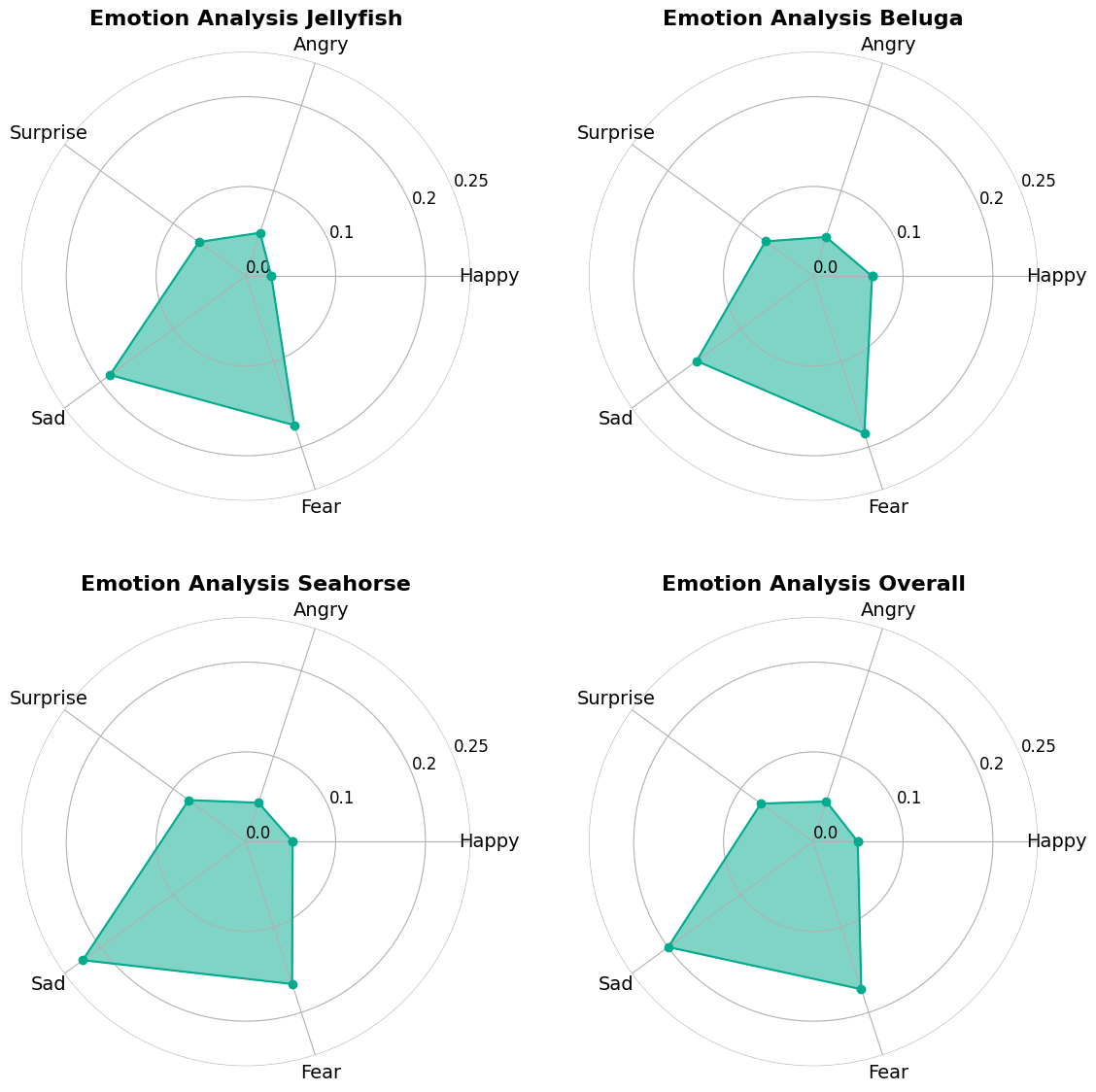}
    \caption{Emotions identified in the chat conversations through text2emotion}
    \label{fig:text2emotion}
\end{figure}

Respondents frequently employed empathetic language, offering apologies and expressions of commiseration. Statements such as '' I'm sorry to hear about how we stupid humans have been affecting your ability to live your life.'' and ''It makes me feel very bad.'' underscored the deep emotional ties participants felt towards the environment. Beyond sadness, there was a collective sense of responsibility and moral duty to act, as many articulated that their personal actions were intertwined with broader environmental well-being. The emotional narratives underscored a strong commitment to the idea that preserving the natural world is integral to both personal and community identity. This emotional connection motivated many participants to seek further positive change, demonstrating the role of emotional engagement as a catalyst for pro-environmental behavior.

\subsubsection{Socioeconomic \& Personal Constraints and Behavioral Trade-offs}
Participants described practical barriers that hinder sustainable lifestyles. Many pointed out that economic limitations—such as the higher cost of eco-friendly alternatives—often force them to choose less sustainable options, with comments like “its hard if i don't have enough money to afford more better options” In addition to financial constraints, some responses highlighted personal challenges, such as limited access to sustainable products, time constraints, or mobility issues, which further complicate efforts to live more sustainably. These constraints illustrate the trade-offs individuals face between environmental ideals and practical realities.

\subsubsection{Skepticism about Individual Impact}
While many respondents expressed readiness to adapt their personal habits for environmental benefit, a notable subset conveyed skepticism about the actual impact of individual actions. Several users framed their efforts as ultimately futile—questioning whether personal measures, such as reducing plastic use, could bring about meaningful environmental change. For instance, one comment noted, “I don't know how much me cutting back would make a real difference. It feels like it needs to be top down.” This perspective was often coupled with a tendency to shift responsibility onto larger entities, with many attributing the dominant role in environmental degradation to corporations and governments rather than individuals. In effect, some users externalized blame, distancing themselves from full accountability by asserting that the solution lies in systemic reform or government intervention rather than individual change.

Additionally, a number of responses revealed cognitive dissonance and behavioral ambivalence. Some participants advocated for reduced plastic use while simultaneously engaging in practices that counteracted these intentions—such as driving more or taking actions that appeared inconsistent with their environmental concerns (e.g. "I will carpool with my Nissan GTR, but will race more as well."). Analyzing these contradictions sheds light on the internal conflicts and cognitive dissonance experienced by individuals as they strive to reconcile their environmental ideals with the practicalities and limitations of everyday life.

\section{Discussion}
Our findings reveal the complex interplay between interactive AI characters, marine species representation, and pro-environmental behavior. Through \textit{OceanChat}, we demonstrate both the potential and limitations of using AI-generated marine characters to foster pro-environmental behavior.

\subsection{Character Interaction and Climate Change Engagement}
Our analysis provides valuable insights into the role of interactive marine characters in shaping environmental attitudes and perceptions of behavior. The Conversational Character Narrative condition demonstrated significant advantages in encouraging participants' sustainable product choices ($\beta = 0.342$, $p = 0.031$) and enhancing participants’  pro-environmental intentions ($\beta = 0.173$, $p < 0.001$) compared to static approaches. Those insights combined with the qualitative analysis from the chat conversations underline that \textit{OceanChat}’s interactive dialogue capabilities appear effective in delivering personalized and engaging environmental messages, helping participants reflect and adjust their sustainability-related actions and intentions in ways that make abstract environmental concerns feel more tangible and actionable.
\begin{figure}
    \centering
    \includegraphics[width=1\linewidth]{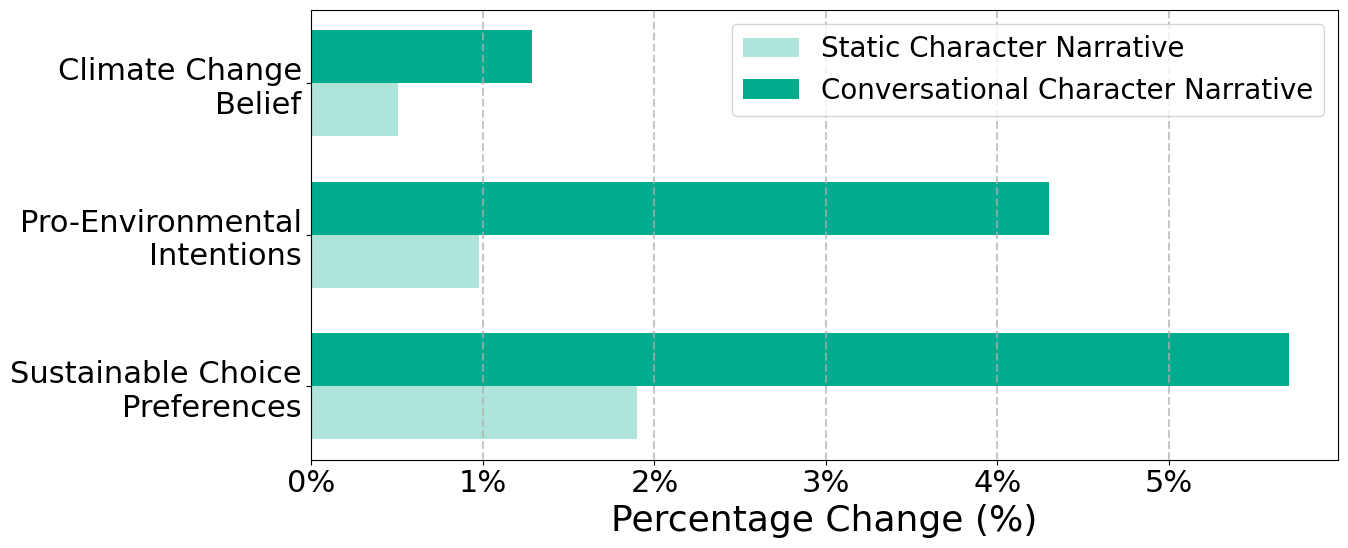}
    \caption{Relative change of the dependent variables approaching significance ($p \leq 0.05$)}
    \label{fig:relative}
\end{figure}

We found that the effectiveness of character interaction varied considerably across different outcome measures. For immediate outcomes like sustainable choice preferences, the interactive condition demonstrated clear benefits. The significant positive effect on sustainable choices suggests that real-time dialogue fosters a stronger connection between abstract environmental messages and participants’ perceived decision-making processes. However, the results revealed key limitations in how these interactions influence deeper psychological and policy-related outcomes. For example, character interactions had no significant effect on psychological distance toward climate change ($\beta = 0.020$, $p = 0.717$) or climate policy adoption ($\beta = 0.315$, $p = 0.535$). These findings suggest that single interactions, while engaging, are insufficient to produce substantive shifts in deeply held beliefs or policy preferences.

Moreover, several qualitative insights reveal that participants often grapple with tangible socioeconomic constraints and cognitive dissonance. Many noted that although they aspire to adopt sustainable practices, factors such as higher costs for eco-friendly alternatives and limited access to sustainable products markedly hinder their ability to follow through. This known disconnect between intention and action emphasizes the need for repeated, context-aware AI interventions that not only inform but also provide practical strategies and local resource guidance to overcome these everyday challenges \cite{ElHaffar2020}.

This highlights the critical distinction between perceived behavior change and actual behavioral outcomes. While character-driven interventions may effectively encourage alignment with sustainability goals and setting intentions, their role should be viewed as one component of a broader strategy for environmental education and advocacy. Integrating repeated interactions, objective measures of behavior, and complementary approaches could help bridge the gap between intention-setting and real-world change.

Interestingly, the interactive condition showed particular strength in fostering sustainable consumption patterns  ($\beta = 0.077$, $p = 0.079$). This suggests that personalized dialogue with marine characters may be especially effective at helping participants reflect on their consumption choices and their environmental impact. The real-time nature of the interaction may create opportunities for marine characters to contextualize environmental issues within participants' daily lives and decision-making processes.

The strong influence of pre-existing attitudes across multiple outcomes presents both challenges and opportunities for interactive character design. While this indicates that character interactions may primarily reinforce existing environmental commitments, it also suggests that carefully designed dialogue systems could help bridge the gap between environmental attitudes and actions. Future iterations of \textit{OceanChat} could potentially incorporate more sophisticated dialogue strategies based on behavioral psychology that help participants transform general environmental concern into specific behavioral commitments \cite{Ajzen1991, Gifford2011b, Gifford2011a, Thaler2008}.

We also observed that the effectiveness of character interaction was mediated by perceptual factors such as anthropomorphism ($\beta = 0.036$, $p = 0.041$) and perceived intelligence ($\beta = 0.043$, $p = 0.082$). This suggests that successful character interaction depends not just on the presence of dialogue, but on creating characters that strike a careful balance between relatable personality and authentic marine species representation. Our findings indicate that when this balance is achieved, interactive marine characters can serve as compelling bridges between human experience and marine conservation issues.

These results extend existing HCI research on persuasive interfaces by demonstrating how AI-generated characters can create personally meaningful environmental narratives through dynamic interaction. While previous work has explored the use of static characters or simple feedback systems, our findings suggest that conversational AI offers new opportunities for creating environmental messages that adapt to individual participants while maintaining consistent character personality and marine species authenticity.

\subsection*{Species Selection and Emotional Connection}
Our research revealed significant differences in how participants emotionally connected with different marine species through AI-generated characters, offering important insights for the future design of marine conservation interfaces. The systematic variation in response across species suggests that character selection critically influences intervention effectiveness. Notably, the jellyfish was found to be perceived as more anthropomorphic compared to the beluga and seahorse conditions (beluga: $\beta = -0.441$, $p < 0.001$; seahorse: $\beta = -0.456$, $p < 0.001$). This implication was unexpected since the jellyfish could be described as more alien and distinct from human characteristics, given its body shape and limited animation variability.

Despite the lower perceived anthropomorphism of belugas and seahorses compared to the jellyfish, the beluga condition showed stronger emotional engagement across multiple measures compared to the seahorse condition. Specifically, perceptions of intelligence ($\beta = 0.280$, $p = 0.001$),  empathetic responses ($\beta = 0.184$, $p = 0.040$), as well as likeability ($\beta = 0.360$, $p < 0.001$) were significantly higher for the beluga condition compared to the jellyfish baseline. These findings align with previous research on charismatic megafauna in conservation, highlighting that certain species can foster deep emotional connections even when they are less anthropomorphic.

However, this apparent advantage of marine mammals presents a design challenge. While leveraging existing emotional connections to cetaceans may increase immediate engagement, it could potentially reinforce problematic hierarchies in marine conservation that prioritize certain species over others. Our results suggest that interaction designers must carefully consider how to harness the emotional appeal of marine mammals while avoiding the perpetuation of species bias.

An unexpected finding emerged regarding the jellyfish condition's effectiveness in communicating climate change impacts. Compared to the seahorse condition, which showed significantly lower impact awareness ($\beta = -0.233$, $p = 0.007$), the jellyfish proved more effective at conveying environmental vulnerability. This challenges conventional assumptions about which species best convey climate change messages and suggests that less anthropomorphic and less animated species like jellyfish may offer unique advantages for environmental communication by emphasizing the distinct and vulnerable nature of marine ecosystems.

The jellyfish's effectiveness may as well stem from its alien nature—its fundamental difference from human experience might actually enhance its ability to communicate the profound changes occurring in marine ecosystems. This finding suggests that future designs should consider how to leverage the "otherness" of marine species rather than always defaulting to highly anthropomorphic representations. Additionally, our analysis revealed interesting patterns in perceived safety across species, with the beluga condition showing higher perceived safety ($\beta = 0.169$, $p = 0.020$) compared to both jellyfish and seahorse conditions. This safety perception correlated with increased willingness to engage with the character's environmental messages, suggesting that perceived threat levels influence the effectiveness of species-based interventions.

These findings have several important implications for the design of marine conservation interfaces. First, they suggest that species selection should be strategically aligned with communication goals. While marine mammals may excel at fostering emotional connection through attributes like intelligence and empathy, other species might better serve specific educational or behavioral objectives by leveraging their unique characteristics. Second, our results indicate that effective character design might benefit from a hybrid approach. Rather than simply maximizing or minimizing anthropomorphic features, designers should consider how to combine the emotional accessibility of marine mammals with the unique communicative advantages demonstrated by less familiar species like jellyfish. Third, the varying effectiveness of different species in conveying climate change impacts suggests that multi-species approaches might be most effective for comprehensive environmental education. Future iterations of \textit{OceanChat} could potentially incorporate multiple marine characters, each leveraging their specific strengths for different aspects of environmental communication.

These insights extend beyond marine conservation to broader questions of how to represent non-human perspectives in interactive systems. Our findings suggest that effective environmental communication through AI characters requires careful consideration not just of technical capabilities, but of the complex relationships between species selection, emotional response, and environmental message effectiveness. By thoughtfully selecting and designing marine species representations, conservation efforts can enhance engagement and foster deeper connections between humans and the marine environment.

\subsection*{Design Implications for AI-Generated Environmental Characters}

Our research findings point to several critical design considerations for developing effective AI-generated characters for environmental communication and behavioral change. These implications span interaction design, emotional engagement, and the broader implementation of AI characters in environmental education.

\subsubsection*{Balancing Anthropomorphism and Authenticity}
The strong influence of both perceived animacy ($\beta = 0.142$, $p = 0.075$) and anthropomorphism ($\beta = 0.174$, $p = 0.007$) on sustainable choice preferences reveals a fundamental design tension in creating marine characters. While anthropomorphic features help foster connection, our results suggest that maintaining authentic marine characteristics is equally crucial for intervention effectiveness. This finding extends beyond simple character design to the core challenge of representing non-human perspectives through AI.

Future systems should consider implementing what we term "graduated anthropomorphism" — where human-like qualities are selectively applied based on interaction context and communication goals. Characters might employ more anthropomorphic features when building initial rapport but shift to more species-authentic representations when discussing specific environmental impacts. This approach could help balance accessibility with educational authenticity.

\subsubsection*{Emotional Design and Behavioral Change}
The significant role of empathy ($\beta = 0.404$, $p < 0.001$) in driving sustainable choices emerged as a crucial design factor. This effect was particularly strong when combined with character likeability ($\beta = 0.783$, $p = 0.026$) in the context of climate change beliefs. These findings suggest that emotional design should be central to character development, not merely an aesthetic consideration.

We recommend implementing emotional scaffolding in character interactions, where emotional engagement is systematically built through a progression of interactions. This progression begins with establishing personal connections, then gradually introduces environmental challenges while integrating specific behavioral change opportunities. The process culminates in reinforcement through shared emotional investment in environmental outcomes.

Qualitative narratives from participant responses vividly underscore this need for emotional scaffolding. For example, many participants expressed genuine sorrow and regret—remarking, ‘I feel so bad about the impact on your home’—which highlights that beyond cognitive appeals, a deep emotional connection drives sustainable intentions. Incorporating these heartfelt narratives into AI dialogue can further enhance engagement by making abstract environmental challenges feel personal and urgent.

\subsubsection*{Interactive Dialogue Design}
The effectiveness of the Conversational Character Narrative condition in promoting pro-environmental intentions ($\beta = 0.173$, $p < 0.001$) highlights the importance of well-designed interactive dialogue. However, the modest effects on information sharing ($R^2 = 0.077$) suggest that current dialogue approaches may not fully capitalize on social diffusion potential.

Future dialogue systems should incorporate contextual resonance—where character responses dynamically adjust to both the immediate interaction context and the participant's demonstrated environmental knowledge and concerns. This involves creating adaptive dialogue paths based on participant engagement levels, integrating local environmental contexts and concerns, and developing personalized behavioral change suggestions. The strategic use of marine species-specific perspectives can further illuminate environmental issues. Furthermore, qualitative feedback underscored the importance of real-world relevance in dialogue. Many participants noted that when the AI’s responses reflected local availability of sustainable products or referenced familiar environmental challenges, they felt a stronger connection. This suggests that adaptive, context-sensitive dialogue not only enhances engagement but also helps translate abstract environmental ideals into actionable, everyday behaviors.

\subsubsection*{Technical Implementation Considerations}
Our findings regarding the impact of the Conversational Character Narrative on perceived intelligence ($\beta = 0.575$, $p < 0.001$) and safety ($\beta = 0.364$, $p < 0.001$) suggest that technical implementation must carefully balance sophistication with approachability. The AI system should demonstrate enough intelligence to maintain credibility while remaining accessible and non-threatening. This approach emphasizes clear communication about the AI's capabilities and limitations while ensuring character responses acknowledge uncertainty when appropriate. Technical sophistication should serve narrative and educational goals rather than being an end in itself. Additionally, the environmental impact of AI implementation should be explicitly considered and minimized.

\subsubsection*{Cross-Platform Integration}
The varying effectiveness of character interactions across different outcome measures suggests the need for integrated, multi-platform approaches. Future designs should consider how AI characters can maintain consistent personality and educational impact across different interaction contexts and technological platforms. This includes developing standardized character frameworks that can be implemented across mobile applications for daily engagement, classroom educational tools, public installation interfaces, social media platforms, and environmental education websites.

\subsubsection*{Scaling and Sustainability}
The strong influence of pre-existing attitudes across multiple outcomes suggests that character-based interventions must be designed with scalability and sustainability in mind. Future systems should incorporate mechanisms for tracking long-term engagement and impact while adapting to changing environmental concerns and scientific understanding. Maintaining character consistency while allowing for evolution and growth remains crucial, as does minimizing computational resources while maximizing environmental benefits. Supporting community building around environmental action represents another key consideration for sustainable implementation.

These design implications provide a framework for developing more effective AI-generated environmental characters while acknowledging the complexities and challenges involved. By carefully considering these elements in future designs, we can create more impactful and sustainable environmental education tools that leverage the unique capabilities of AI while serving crucial conservation goals.

\section*{Environmental Impact of AI-Based Sustainability Interventions}
Our analysis of \textit{OceanChat}’s environmental footprint highlights the carbon implications of utilizing AI for sustainability education. Based on estimates, a single conversation (3 LLM queries + ~30s of TTS usage) with \textit{OceanChat} consumes approximately 0.0016 kWh of energy \cite{baeldung2024chatgpt, Aramon2024}. For context, this is equivalent to 11\% of a smartphone charge (assuming 0.015 kWh per charge). With an estimated user base of 10,000 monthly active users engaging in weekly conversations, our system’s annual energy consumption is approximately 832 kWh. Using an average grid carbon intensity of 0.322 kgCO$_2$e/kWh, this results in approximately 0.27 metric tons of CO$_2$ emissions annually, equating to about 1.8\% of an average American's footprint. \cite{epa2023, iea2023}.

While \textit{OceanChat}’s annual footprint of 0.27 metric tons of CO$_2$e represents a relatively small environmental cost, this must still be justified by measurable behavioral outcomes. Strategies to further reduce energy use—such as response caching, request batching, and a hybrid model system—should be implemented to maintain or reduce this low footprint as the user base scales. For example, deploying smaller models for routine interactions could cut per-conversation energy use by up to 40\%, while renewable-powered data centers exemplify how operational choices amplify efficiency gains. By aligning system design with sustainability principles, future implementation should ensure that even marginal energy costs translate to outsized educational benefits.

Future implementations should consider environmental impact as a core design metric, not an afterthought. This includes exploring emerging efficient AI architectures, implementing strict computation budgets per interaction, and developing clear metrics for balancing system performance with energy consumption. The goal should be to maximize the ratio of behavioral impact to computational cost, ensuring that AI-powered sustainability interventions contribute positively to their intended environmental goals.

\section{Limitations and Future Work}
While our study demonstrates the potential of AI-generated marine characters in facilitating pro-environmental behavior, several limitations should be acknowledged. The strong influence of pre-existing attitudes across multiple outcomes suggests that character-based interventions may be more effective at reinforcing existing environmental commitments than creating new ones. Future work should explore how to better reach and engage individuals with lower initial environmental concern. Additionally, the relatively short-term nature of our study leaves open questions about the durability of behavioral changes induced through character interaction. Longitudinal studies will be crucial for understanding how these effects persist over time and what design features might support sustained engagement with environmental issues. Finally, while we carefully considered the environmental impact of our AI implementation, future work should continue to explore ways to minimize the computational resources required for interactive character experiences while maximizing their environmental benefits.

\section{Conclusion}
\textit{OceanChat} demonstrates both the promise and complexity of using AI-generated marine characters for more pro-environmental behavior. While interactive characters can effectively promote sustainable intention setting, sustainable choices, and emotional connection with marine life, their impact varies significantly across different outcomes and species. These findings contribute to our understanding of how to design effective AI characters while highlighting important considerations around species representation, interaction design, and the limitations of character-based approaches to environmental advocacy. Future work in this area should focus on addressing these limitations while building on the demonstrated potential of interactive marine characters to foster environmental awareness and action.

\bibliographystyle{ACM-Reference-Format}
\bibliography{references}
\end{document}